 \documentclass[sigplan,screen]{acmart}

\copyrightyear{2026}
\acmYear{2026}
\setcopyright{cc}
\setcctype{by}
\acmConference[ASPLOS '26]{Proceedings of the 31st ACM International Conference on Architectural Support for Programming Languages and Operating Systems, Volume 1}{March 22--26, 2026}{Pittsburgh, PA, USA}
\acmBooktitle{Proceedings of the 31st ACM International Conference on Architectural Support for Programming Languages and Operating Systems, Volume 1 (ASPLOS '26), March 22--26, 2026, Pittsburgh, PA, USA}
\acmPrice{}
\acmDOI{10.1145/3760250.3762226}
\acmISBN{979-8-4007-2165-6/26/03}



\AtBeginDocument{%
  }

\usepackage{graphicx}
\usepackage[export]{adjustbox}
\usepackage{graphicx}
\usepackage{mdframed}
\usepackage{textcomp}
\usepackage{pgfplots}
\usepackage{tikz}
\usepackage{pgf-pie}
\usepackage{algpseudocode}
\usepackage{lipsum}
\usepackage[linesnumbered,titlenumbered,ruled,noend]{algorithm2e}
\usepackage{algorithmicx}
\usepackage{multirow}
\usepackage{fancyhdr}
\usepackage[all]{nowidow}
\definecolor{lstgrey}{rgb}{0.95,0.95,0.95}
\usepackage{listings}
\usepackage{textcomp}
\usepackage{svg}
\usepackage{tikzscale}

\usepackage{enumitem}
\usepackage{tabu,multirow,makecell}
\usepackage{pifont}
\usepackage{enumerate}
\usepackage{xurl}
\usepackage[flushleft]{threeparttable}
\usepackage{enumitem}
\usepackage{subcaption}
\usepackage[colorinlistoftodos]{todonotes}
\usepackage{arydshln}
\usepackage{stfloats}

\renewcommand{\textrightarrow}{$\rightarrow$}

\usepackage{etoolbox}
\usepackage{xcolor}
\usepackage{pdfpages} 
\usepackage{listings}

\definecolor{dkgreen}{rgb}{0,0.6,0}
\definecolor{gray}{rgb}{0.5,0.5,0.5}
\definecolor{mauve}{rgb}{0.58,0,0.82}
\definecolor{OliveGreen}{rgb}{0,0.6,0}
\definecolor{mahogany}{rgb}{0.75, 0.25, 0.0}
\definecolor{darkmidnightblue}{rgb}{0.0, 0.2, 0.4}
\definecolor{navyblue}{rgb}{0.0, 0.0, 0.5}
\definecolor{apricot}{rgb}{0.98, 0.81, 0.69}
\definecolor{antiquewhite}{rgb}{0.98, 0.92, 0.84}
\definecolor{brickred}{rgb}{0.8, 0.25, 0.33}
\definecolor{bananamania}{rgb}{0.98, 0.91, 0.71}
\definecolor{bisque}{rgb}{1.0, 0.89, 0.77}
\definecolor{champagne}{rgb}{0.97, 0.91, 0.81}
\definecolor{eggshell}{rgb}{0.94, 0.92, 0.84}
\definecolor{darkolivegreen}{rgb}{0.33, 0.42, 0.18}
\definecolor{phthalogreen}{rgb}{0.07, 0.21, 0.14}
\definecolor{richblack}{rgb}{0.0, 0.25, 0.25}
\definecolor{anti-flashwhite}{rgb}{0.95, 0.95, 0.96}

\definecolor{pie1}{gray}{0.2}
\definecolor{pie2}{gray}{0.4}
\definecolor{pie3}{gray}{0.6}
\definecolor{pie4}{gray}{0.8}
\definecolor{pie5}{gray}{0.9}

\definecolor{pie11}{gray}{0.1}
\definecolor{pie12}{gray}{0.25}
\definecolor{pie13}{gray}{0.4}
\definecolor{pie14}{gray}{0.55}
\definecolor{pie15}{gray}{0.70}
\definecolor{pie16}{gray}{0.85}

\definecolor{pastel11}{HTML}{FFEEF4}
\definecolor{pastel12}{HTML}{E4E4D0}
\definecolor{pastel13}{HTML}{AEC3AE}
\definecolor{pastel14}{HTML}{94A684}
\definecolor{pastelRed}{HTML}{FF90BC}

\definecolor{pastel41}{HTML}{A1CCD1}
\definecolor{pastel42}{HTML}{F4F2DE}
\definecolor{pastel43}{HTML}{E9B384}
\definecolor{pastel44}{HTML}{7C9D96}

\definecolor{color11}{HTML}{fee8c8}
\definecolor{color12}{HTML}{fdbb84}
\definecolor{color13}{HTML}{e34a33}

\definecolor{pastel31}{HTML}{f1eef6}
\definecolor{pastel32}{HTML}{d7b5d8}
\definecolor{pastel33}{HTML}{df65b0}
\definecolor{pastel34}{HTML}{ce1256}

\definecolor{pastel51}{HTML}{a6611a}
\definecolor{pastel52}{HTML}{dfc27d}
\definecolor{pastel53}{HTML}{f5f5f5}
\definecolor{pastel54}{HTML}{80cdc1}
\definecolor{pastel55}{HTML}{018571}

\newcommand{\shephered}[1]{\textcolor{black}{#1}}

\newenvironment{shepherding}{\par\color{black}}{\par}

\usetikzlibrary{shadows,patterns,shapes,arrows,decorations.pathmorphing,backgrounds,positioning,fit,plotmarks,calc,spy}
\usepgfplotslibrary{patchplots}
\definecolor{darktan}{rgb}{0.57, 0.51, 0.32}
\definecolor{desert}{rgb}{0.76, 0.6, 0.42}
\definecolor{desertsand}{rgb}{0.93, 0.79, 0.69}
\pgfplotsset{compat=1.5,
	/pgfplots/ybar legend/.style={
		/pgfplots/legend image code/.code={%
			\draw[##1,/tikz/.cd,bar width=3pt,yshift=-0.2em,bar shift=0pt]
			plot coordinates {(0cm,0.8em)};},
	},
}

\newcommand{\refauthors}{Zhenyu Bai, Pranav Dangi, Rohan Juneja, Zhaoying Li, Zhanglu Yan, Huiying Lan, and Tulika Mitra}

\begin{document}

\title{A Data-Driven Dynamic Execution Orchestration Architecture}

\settopmatter{authorsperrow=4}

\author{~}
\affiliation{%
\institution{~}
\state{~}
\country{~}
}

\author{Zhenyu Bai}
\authornote{Both authors contributed equally to this research.}
\email{zhenyu.bai@nus.edu.sg}
\affiliation{
    \department{School of Computing}
    \institution{National University of }
    \city{Singapore}
    \country{Singapore}
}

\author{Pranav Dangi}
\authornotemark[1]
\email{dangi@comp.nus.edu.sg}
\affiliation{
    \department{School of Computing}
    \institution{National University of }
    \city{Singapore}
    \country{Singapore}
}

\author{~}
\affiliation{%
\institution{~}
\state{~}
\country{~}
}

\author{Rohan Juneja}
\email{rohan@comp.nus.edu.sg}
\affiliation{
    \department{School of Computing}
    \institution{National University of }
    \city{Singapore}
    \country{Singapore}
}

\author{Zhaoying Li} 
\authornote{Corresponding author}
\email{zhaoying@comp.nus.edu.sg}
\affiliation{
    \department{School of Computing}
    \institution{National University of }
    \city{Singapore}
    \country{Singapore}
}

\author{Zhanglu Yan}
\email{zlyan@nus.edu.sg}
\affiliation{
    \department{School of Computing}
    \institution{National University of }
    \city{Singapore}
    \country{Singapore}
}

\author{Huiying Lan}
\email{huiying.lan@lumai.co.uk}
\affiliation{
  \institution{Lumai Ltd.}
  \city {Oxford}
  \country{UK}
}

\author{Tulika Mitra} 
\email{tulika@comp.nus.edu.sg}
\affiliation{
    \department{School of Computing}
    \institution{National University of }
    \city{Singapore}
    \country{Singapore}
}


\renewcommand{\shortauthors}{Zhenyu Bai et al.}
\newcommand\name{\textit{Canon}}

\begin{abstract} 
Domain-specific accelerators deliver exceptional performance on their target workloads through fabrication-time orchestrated datapaths. However, such specialized architectures often exhibit performance fragility when exposed to new kernels or irregular input patterns. In contrast, programmable architectures like FPGAs, CGRAs, and GPUs rely on compile-time orchestration to support a broader range of applications; but they are typically less efficient under irregular or sparse data. Pushing the boundaries of programmable architectures requires designs that can achieve  efficiency and high-performance on par with specialized accelerators while retaining the agility of general-purpose architectures.

We introduce {\name}, a parallel architecture that bridges the gap between specialized and general purpose architectures.  {\name} exploits data-level and instruction-level parallelism through its novel design. First, it employs a novel dynamic data-driven orchestration mechanism using programmable Finite State Machines (FSMs). These FSMs are programmed at compile time to encode high-level dataflow per state and translate incoming meta-information (e.g., sparse coordinates) into control instructions at runtime. Second, {\name} introduces a time-lapsed SIMD execution in which instructions are issued across a row of processing elements over several cycles, creating a staggered pipelined execution. These innovations amortize control overhead, allowing dynamic instruction changes while constructing a continuously evolving dataflow that maximizes parallelism. Experimental evaluation shows that {\name} delivers high performance across diverse data-agnostic and data-driven kernels while achieving efficiency comparable to specialized accelerators, yet retaining the flexibility of a general-purpose architecture.

\end{abstract}

\maketitle

\section{Introduction}
\label{sec:intro}

An ideal processor would allocate nearly all its resources to computation while minimizing the cost of control and data movement. However, the end of Dennard scaling, the slowdown of Moore’s law, and the disparate scaling of memory relative to logic have constrained the performance and efficiency gains of classical von Neumann architectures, moving them away from the ideal processor envisioned~\cite{horowitzmemory, diag}. As these architectures increasingly devote resources to control and data movement, their computational efficiency decreases. Against this backdrop, the rise of domain-specific compute-intensive applications has triggered a Cambrian explosion of novel, non–von Neumann accelerator architectures. 

These domain-specific architectures incorporate specialized compute units and manage data dependencies through hardwired datapaths that \textit{mimic} the application's intrinsic data flow~\cite{mit-dataflow}. This hardwiring reduces the cost of control and memory access relative to the computation. Essentially, these architectures rely on \textit{the fabrication-time orchestration} or configuration of the compute and data dependencies. Examples include Tensor Processing Units (TPUs)\cite{tpu}, AI accelerators\cite{eyeriss, vitcod, sanger}, media and protocol accelerators~\cite{protobuf}, and sparse tensor accelerators~\cite{flexagon, sigma, gamma, zed}, each designed solely for the respective workloads. However, when tasked with workloads beyond their intended domain, these architectures exhibit extreme \textit{fragility} in performance~\cite{trips}. 

In contrast, programmable architectures, such as Coarse-Grained Reconfigurable Arrays (CGRAs) and FPGAs, rely on sophisticated mapping, placement, and routing software to orchestrate computation on their reconfigurable fabrics. By spatially distributing computation and communication at compile time, they achieve flexibility compared to specialized accelerators. 
GPUs, on the other hand, leverage an execution model with massive thread-level parallelism, relying on compilers and schedulers to manage computation. While they offer more adaptability than fixed-function accelerators, their efficiency hinges on structured and regular workload decomposition and inherent parallelism. 

Irregularity is becoming an increasingly critical factor in modern workloads. For instance, sparse tensor operations in machine learning (ML) workloads bring irregularities with wide-ranging sparsity from 5\% to 95\%, and can appear in structured or unstructured forms, either known at compile time or determined at runtime. Supporting a broad range of sparse kernels has become essential~\cite{deepseek-nsa, longformer, bigbird}. 
Ultimately, GPUs, FPGAs, and CGRAs perform well on regular or data-parallel tasks but are less efficient for workloads with irregular dataflow and memory access patterns. While these architectures generally handle workload variations better than specialized accelerators, the compile-time orchestration and reliance on massive thread level parallelism can lead to performance degradation when faced with dynamic or unpredictable data patterns. This emphasises the need for a hybrid approach that integrates static and dynamic decision-making to handle an assortment of regular and irregular workloads with minimal resource overhead.

\paragraph*{Contributions:}
We propose {\name}\footnote{\textit{Canon} in music stands for imitation with an offset, or a guiding principle.}, a novel parallel architecture that transcends traditional specialization-flexibility tradeoffs.
Our objective is to push the boundaries of programmable architectures, aligning with the idea that extreme domain specialization may often be superfluous~\cite{specialization-question}. 

\begin{figure}
    \centering
    \includegraphics[width=.95\linewidth]{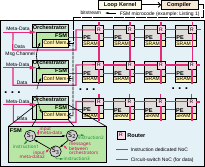}
    \caption{{\name} Hardware Architecture}
    \label{fig:arch-framework}
\end{figure}

As shown in Figure~\ref{fig:arch-framework}, {\name} is based on a 2D-mesh spatial architecture composed of Processing Elements (PEs). Unlike conventional reconfigurable architectures that rely exclusively on compile-time orchestration, {\name} exploits a two-tier approach. {\bf {\name} first leverages inherent workload regularities, specifically, data-level and instruction-level parallelism, to schedule predictable high-level dataflows; and then uses dynamic scheduling in hardware via a lightweight programmable orchestrator to handle irregularities.} Therefore, the cost of dynamic scheduling and control is minimized by confining it to the irregular aspects of the workload (e.g., arbitrary input patterns), while the bulk of the execution benefits from the efficiency of regular dataflows. The orchestrator incorporates a FSM which acts as an on-the-fly data-to-instruction translator. This FSM is programmed by the compiler, which relies on static analysis of the high-level dataflow and compute organization; and at runtime, the FSM relies on input-data and neighbor messages as triggers for generating instructions. 

Second, {\name} incorporates a time-lapsed SIMD execution combined with distributed memories, where instructions propagate through PEs over multiple cycles, amortizing control cost and enhancing scalability. While different PEs may be at various stages of execution at an instant, they eventually perform the same operation over time. Consequently, \textbf{{\name} exploits an evolving dataflow across the fabric, where execution patterns are constructed, switched, and adapted over time at fine granularity to suit an unpredictable input without sacrificing parallelism.}

Control and synchronization are completely abstracted from the compute units in the PE array by resorting to orchestrators and issuing instructions in a time-lapsed manner. This abstraction enables \textit{a priori} look-ahead into decisions, yielding predictable and deterministic behavior across the fabric, even under dynamic orchestration. The hardware sustains high parallelism, maximizes utilization, balances workload efficiently, and ultimately achieves consistently high throughput across diverse kernels and input data patterns.
\textbf{{\name} matches the performance of dense, unstructured and structured sparse accelerators in their own specialization domains with minimal efficiency degradation while also supporting a broad array of other parallel workloads typically handled by reconfigurable architectures.} To our knowledge, {\name} is the first architecture to sustain such high performance across such a diverse set of data-agnostic and data-driven kernels.

\section{{\name} Architecture}

$\name$ is designed to effectively handle both regular workloads and data irregularities, with the primary intention of ensuring a stable performance and minimal fragility. 
This section provides an overview of the architecture's design principles that combine dynamic orchestration with a scalable reconfigurable fabric composed of PEs in a 2D mesh topology with a time-lapsed SIMD execution. Each PE has a vector lane for computation, associated registers, a router for communication with neighboring PEs, and a local data memory. The detailed micro-architectural implementation and mapping are further discussed in later sections.

\label{sec:staggered-inst-issue}

\begin{figure}
    \centering
    \includegraphics[width=.75\linewidth]{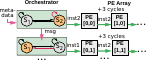}
    \caption{Orchestration and Instruction issue in {\name}}
    \label{fig:data-driven}
\end{figure}

\paragraph{Data-Driven Orchestration:}
Figure~\ref{fig:data-driven} illustrates the mechanism where each row of processing elements is managed by a dedicated orchestrator, which is a lightweight, programmable finite-state machine. A bitstream configures the FSM with states representing high-level operational modes (e.g., \shephered{computing multiplication}, sending data to neighborhood PEs, or accumulating partial sums). The FSM dynamically issues instructions to its row of PEs reacting to input meta-data (e.g., sparse coordinates) and messages from neighbors \shephered{if required}. For instance, in Figure~\ref{fig:data-driven}, both the rows are initially issued instruction 2 $(inst2)$ corresponding to State $S_2$; subsequently, a message from the orchestrator at the North to South triggers the second row’s transition from State $S_2$ to $S_1$, changing its output instruction from $inst2$ to $inst1$. This dynamic orchestration, which is a hybrid of compile-time mapping with runtime decision making, enables the architecture to adapt seamlessly to both regular and irregular input data patterns.

\begin{figure}[h]
    \centering
    \includegraphics[width=\linewidth]{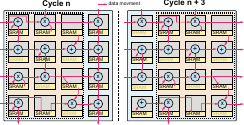}
    \caption{Time-lapsed SIMD with staggered instruction issue}
    \label{fig:example-exec}
\end{figure}

\paragraph{Time-Lapsed SIMD Execution:}
As depicted in Figure~\ref{fig:example-exec}, the architecture employs a time-lapsed SIMD execution model wherein instructions generated by the FSM-based orchestrator propagate through the PE array over multiple cycles via a dedicated instruction network in a staggered manner. Unlike conventional SIMD execution, where a single instruction is broadcast simultaneously to all PEs, here an instruction such as “multiply inputs from SRAM and East; Send Output South” is issued to the first PE in cycle 1, then traverses a 3-cycle pipeline before reaching the second PE in cycle 4. This staggered instruction issue ensures that while different PEs can behave differently at a given timestamp, they ultimately execute an identical sequence of operations on their respective data. The behavior, including NoC and memory actions, is replicated across the fabric. For instance, in Figure~\ref{fig:example-exec}, the first columns' compute, memory, and NoC behavior are recreated three cycles later by the next columns.
\subsection{\textbf{Reduced Control Cost}}
An important issue with traditional von Neumann and many reconfigurable architectures is the high overhead of control logic~\cite{diag, plaid}. In \name, orchestrators provide a lightweight control mechanism, reducing the per-PE control burden. They generate and distribute control signals via an instruction-dedicated NoC from the periphery of a PE row through staggered instruction issue. 
By offloading control and instruction distribution, we preserve a lightweight compute fabric, enhance scalability, simplify the programming model, and achieve high energy efficiency.

\textbf{Synchronization:} Each cycle, the orchestrator dispatches an instruction to the first PE of the row, which executes it and subsequently passes it along to the next PE. Consequently, every PE in a row ultimately performs the same instruction albeit at staggered cycles and on distinct data. Each PE operates with a fixed pipeline latency of 3 cycles, ensuring that the sequence of actions initiated by the first PE (including data exchanges and memory accesses) is consistently replicated by subsequent PEs with a delay corresponding to the instruction propagation. This predictable behavior enables us to abstract inter-PE synchronization through orchestrator-level coordination. This also means effectively enhancing the compute-to-control ratio, maximizing computational density, and reducing control overhead.


\textbf{Abstracted \& Deterministic Irregularity Handling:}
The orchestrator is primarily designed to manage dynamic input variations, such as sparsity, by effectively addressing the challenges posed by irregular workloads and dynamic data dependencies. While CPUs and GPUs resolve such dependencies through memories, dataflow architectures depend largely on the NoC for data movement. In regular applications with predictable communication patterns, static routing via circuit-switched or hardwired NoCs is effective. However, for complex applications characterized by irregularities and variable data dependencies, packet-switched or dynamic NoCs are preferable due to their adaptability, despite incurring additional hardware overhead from mechanisms such as backpressure and virtual channels~\cite{noc, jerger2017chip, scalable_interconnects}.

We design a \textit{dynamically managed circuit-switching} scheme that handles both regular and irregular workloads with minimal overhead. Thanks to deterministic timing across the PE array from abstracted synchronization, runtime control and congestion management within the PE’s NoC become unnecessary. The orchestrators manage irregularities externally, using their insight into PE determinism and input data patterns to make dynamic decisions. These decisions are embedded in the instruction stream, ensuring that the circuit-switched configurations accurately reflect the required data dependencies during execution.

\subsection{Reduced Data-Handling Costs}

{\name} employs a distributed memory system in which each PE features local memory and communicates via the NoC. The orchestrators dynamically configure both the NoC and the memory through instructions, to efficiently support regular and irregular workloads. Although \shephered{circuit-switch} NoCs are highly efficient in data movement~\cite{jerger2017chip, horowitzmemory}, their fixed topology and limited bandwidth constrain their ability to manage complex dependencies, necessitating reliance on memory.

{\name} prioritizes the NoC for mapping inherently regular compute patterns and data transfers, while it opportunistically resorts to memory when handling irregularity. In this design, memory serves a dual purpose and is partitioned respectively into two segments to prevent port saturation and minimize fragmentation: a larger segment for static data (e.g., ML weights) and a smaller scratchpad that serves as a read/write buffer to handle complex dependencies on the fly. Both segments support single-cycle random access.

As a \emph{storage} device, the static memory \shephered{(named \emph{data memory} in the figures and texts below for simplicity)} is mainly used for input and output data, enabling effective data reuse, reducing off-chip bandwidth requirements, and accommodating random accesses, particularly beneficial for irregular workloads with runtime-determined patterns. For resolving \shephered{complex runtime} data dependencies, the scratchpad holds intermediate data during execution, akin to \shephered{the register file of the} von Neumann architectures \shephered{such as GPUs and CPUs}, or acts as a buffer to amortize runtime irregularities\shephered{, the way we use it in sparse computation as detailed later}. 
{\name} decouples local memory management from the compute units by integrating both software and hardware control, drawing inspiration from Explicitly Decoupled Data Orchestration (EDDO) architectures~\cite{buffets}. The compiler directs data distribution and programs kernel-specific data management policies into the orchestrator FSM, enabling dynamic memory management based on orchestrator instructions.

\section{CANON Micro-architecture}
\label{sec:micro-architecture}
\begin{figure}[h]
    \centering
    \includegraphics[width=\linewidth]{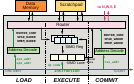}
    \caption{Architecture of 3-stage {\em CANON} PE pipeline} 
    \label{fig:PE-arch}
\end{figure}

The micro-architecture of the PE and the orchestrator is illustrated in Figure ~\ref{fig:PE-arch} and Figure~\ref{fig:orchestrator-arch}, respectively. Each PE functions as a 3-stage pipeline. 
At \lstinline{LOAD} stage, data is loaded from the scratchpad memory, data memory, or the NoC into the vector-lane registers for input operands. The \lstinline{COMPUTE} stage performs computations using a vector lane that processes four words in parallel. The results are written to the scratchpad, registers (e.g. for accumulation), data memory or sent to neighborhood PEs through the NoC at the \lstinline{COMMIT} stage. 

\subsection{ISA and PE Control}
PEs do not include complex control logic; they primarily rely on orchestrator-issued instructions. These instructions are streamed through the PE pipeline, dictating their execution behavior, i.e., the router, compute, and memory behavior. The instruction format is straightforward, specifying the operation, operand addresses, and destination addresses:

\begin{lstlisting}[basicstyle=\footnotesize\ttfamily]
        <inst> ::= <op> <op1_addr> <op2_addr> <res_addr>
\end{lstlisting}

To simplify the instruction format, the scratchpad, data memory, router, and SIMD registers share a unified address space. The specific memory accessed or NoC switching action is inferred from the address.

Since instructions are executed in a pipelined manner, hardware resource contention is minimized.
The read ports of the data memory and scratchpad are accessed only during the \lstinline{LOAD} stage if \lstinline{op1_addr} or \lstinline{op2_addr} corresponds to these memories, while write ports are used exclusively during the \lstinline{COMMIT} stage if \lstinline{res_addr} targets them.
The NoC switch is active in both \lstinline{LOAD} and \lstinline{COMMIT} stages when transferring data. Due to router constraints, it supports only one data transfer per cycle per direction. 
As pipeline stages are executed in parallel (ILP), a single instruction cannot simultaneously read from and write to the same NoC direction to prevent conflict with other instructions executed in parallel. This restriction is enforced at compile time.

\subsection{Orchestrator Microarchitecture}
\begin{figure}[h]
    \centering
    \includegraphics[width=\linewidth]{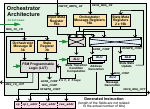}
    \caption{Architecture of the programmable Orchestrator}
    \label{fig:orchestrator-arch}
\end{figure}

The orchestrator is designed to generate instructions at runtime using its FSM, which serves as a data-to-instruction translation function. Its primary roles are to produce instruction fields, update its internal state, and send messages to neighboring orchestrators. As shown in Figure~\ref{fig:orchestrator-arch}, the FSM's internal state is maintained using two types of registers: the \textit{State Register}, which holds the current state, and the \textit{State Meta Registers}, storing value-based state information such as iteration counts or memory status, \shephered{defined by the compiler} depending on the target kernel. 

The state transition function is defined at compile time, while the state transitions are triggered at runtime by external events, including metadata from the input stream captured in the \textit{Input Meta Register} and messages from neighboring orchestrators in the \textit{Orchestrator Message Register} (for content) and the \textit{Orchestrator Message ID} (for message type). 
We note that the semantics of the states and messages are not fixed by the hardware but instead defined by the compiler along with the definition of state transition logic.

The registers are processed for four key functions: computing the state transition condition, generating addresses, calculating the new state, and generating messages. The condition computation is statically configured (all gray components in the figure), while the other three functions are dynamically configured (blue components) by programmable logic depending on the FSM state. This programmable logic is a lookup table (LUT) unit capable of implementing any combinational logic function of its inputs. The LUT is implemented as SRAM. It contains $2^{10}$ entries, corresponding to all possible configurations of its 10 input bits ($2^{3+3+2\times2}$). Each entry outputs 48 bits for configuring the dynamic components, resulting in 6 KB SRAM.
The dynamic control logic is defined by the programmer or compiler by specifying the encoding of the state, state meta, orchestrator messages, and message IDs. Before kernel execution, this data-instruction translation logic is prefilled into the LUT as a `bitstream', enabling programmability of the orchestrator.

\section{{\name} Application Mapping}

As {\name} eliminates much of the control from PEs and has a new execution paradigm, it is essential to demonstrate the architecture's programmability and generalizability across various kernels. 

\begin{shepherding}
Exploiting irregularity embedded in otherwise regular workloads is fundamental to {\name}’s design. {\name}’s compilation framework targets nested loops with conditional computation. Figure~\ref{fig:compilation-framework} outlines the flow. The frontend recovers loop semantics via polyhedral analysis of array accesses, inter-iteration dependences, and the impact of conditional (predicated) execution.

These analyses drive a second stage of spatial/dataflow exploration, where the compiler evaluates how iterations should be placed on the fabric. Different loop semantics induce different control and data-movement patterns; under a fixed fabric topology, many candidate mappings prove illegal or clearly suboptimal. The resulting search space is vast, and finding globally optimal mappings and placements on {\name} remains an open problem. Currently, the workflow combines static loop analyses with human intervention to determine a final dataflow.

\begin{figure}[h]
        \centering
        \includegraphics[width=\linewidth]{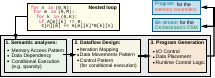}
        \caption{{\name} Compilation Framework}
        \label{fig:compilation-framework}
\end{figure}

Once the spatial dataflow is fixed, the data and basic possible computations to be mapped at each PE are known. We encode this as three intermediate representations: (i) data provision/I/O control from higher-level memories to the PE array, (ii) the initial data placement across the array, and (iii) the control logic. Because {\name} follows the EDDO schema~\cite{buffets}, the first two IRs drive code generation for the asynchronous memory movers around the array, ensuring each PE receives its data at the right time. The compute/control schedule is emitted as FSM microcode (cf. Listing~\ref{lst:spmm-fsm}), which is finally compiled into the FSM bitstreams.

Compiling high-performance binaries for arbitrary sparse workloads remains an open problem and is critical to {\name}’s efficiency. Our end-to-end framework, especially the front-end from high-level code (e.g., PyTorch, C), is still under development. In this work, we instead expose the underlying mapping logic and demonstrate it on representative sparse tensor kernels.
\end{shepherding}


\subsection{\textbf{Mapping Sparse Kernels}}
Sparse kernels, particularly sparse tensor operations, are among the most popular and representative HPC workloads that involve input-dependent control flow. These operations create bottlenecks in applications spanning high-performance computing (HPC) and machine learning~\cite{gamma, sigma, flexagon, deepseek-nsa, deepseek-v3}. Moreover, these workloads involve varying sparse operations with different degrees of randomness or structure in inputs. Sparse tensor operations highlight two primary challenges common to applications with irregular input data. First, the sparsity in inputs introduces irregular memory access patterns, leading to memory bottlenecks. Second, the uneven distribution of non-zero elements results in workload imbalance across compute units, necessitating specialized hardware for load balancing and synchronization. Traditional architectures, including GPGPUs~\cite{tesnorcore} and systolic arrays~\cite{tpu}, often suffer from under-utilization of compute units due to this. 
We demonstrate the mapping of various dense and sparse kernels on {\name} hardware.


\begin{shepherding}
\subsubsection{\textbf{Case Study: SpMM}} \label{sec:spmm}

Figure~\ref{fig:spmm-sddmm-dataflow}(a) illustrates the SpMM dataflow used in our design, derived from Gustavson’s algorithm~\cite{gustavson, gamma, zed, flexagon}. In this approach, rows of the sparse matrix $A$ are independently processed in parallel to generate the corresponding output rows. The non-zero entries of $A$ are streamed into the corresponding rows of the PE array.

Each PE holds a tile of the dense matrix $B$ in its local memory. PEs aligned along the same row of the array store identical rows of $B$, but are partitioned across distinct column segments. For every non-zero element $a_{ij}$ in a given row of $A$, the row of $B$ indexed by $j$ is fetched from local memory. This enables the PE to perform scalar-vector multiplications between $a_{ij}$ and the local portions of $B$, generating partial vector products (\lstinline{psums}).

These psums are then propagated vertically along the columns of the array, accumulating partial results as they traverse downstream PEs. The final accumulated values exit the bottommost PE in each column, thereby producing the corresponding entries of the output matrix $C$. The complete pseudocode for this dataflow, derived by tiling and reordering the canonical triple-for-loop SpMM kernel, is provided in Listing~\ref{lst:spmm-code} in the Appendix.

This mapping delivers high efficiency on our architecture by simultaneously tackling three key challenges: \textbf{irregular access patterns}, \textbf{reduction-induced dependencies}, and \textbf{workload imbalance}, while preserving fine-grained parallelism across the array.

\begin{figure}
\begin{subfigure}{.63\linewidth}
    \centering
    \includegraphics[width=\linewidth]{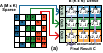}
\end{subfigure}
\begin{subfigure}{.36\linewidth}
    \centering
    \includegraphics[width=\linewidth]{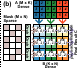}
\end{subfigure}
\caption{(a) SpMM and (b) SDDMM dataflow mapping}
    \label{fig:spmm-sddmm-dataflow}
\end{figure}


\textbf{Local Random Accesses.}
We tile and partition the dense matrix $B$ such that all PEs within the same physical row store identical rows of $B$, each holding a distinct column segment. This layout ensures that, for any incoming non-zero element from $A$, all PEs in the row access the same address offsets in their respective local memories. As a result, local memory accesses remain uniform and fully deterministic, aligning with {\name}'s staggered instruction issue model. Since all accesses to $B$ are serviced from each PE’s local memory, this design eliminates contention and scales efficiently with increasing PE array sizes, bypassing the bandwidth bottlenecks typically associated with shared memory architectures.

\begin{figure}[h]
    \centering
    \includegraphics[width=0.8\linewidth]{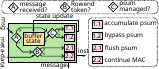}
    \caption{SpMM decision tree. Condition A checks for messages from neighbour orchestrators, Condition B checks for end of row, Condition C checks local context window of partial sums}
    \label{fig:fsm-decision}
\end{figure}


\textbf{Load Balancing via Asynchronous Reduction.}
When the reduction dimension ($K$) is spatially mapped across rows, naive accumulation strategies introduce write-after-write (WAW) hazards, particularly in the presence of sparse and irregular $A$ matrices. In such cases, PEs responsible for rows with higher non-zero densities become stragglers, causing downstream stalls and underutilization.

{\name} mitigates this through \textit{asynchronous reduction}, enabled by the associativity of addition. Each PE concurrently has two roles: performing MAC operations for its local non-zero elements and accumulating partial sums propagated from upstream PEs. Upon completing MACs for a row, the PE advances to copmuting the MACs of next row, independent of whether upstream PEs are straggling or in advance. It only switches to partial sum accumulation when the upstream PE sends it a message for the same.

This decision is made dynamically by the orchestrator using the FSM in Figure~\ref{fig:fsm-decision}. When a control message arrives from the north (condition A = yes), indicating that upstream PEs are flushing partial sums, the orchestrator switches the local PEs to handle accumulation (paths 1.1 or 1.2). If no such message is received, the orchestrator allows PEs to proceed with local MAC computation (path 2.2). This decoupled and dynamic scheduling ensures forward progress even under load imbalance, thereby preserving high array utilization.

\textbf{Load Balancing via Explicit Buffer Management.}
While asynchronous reduction alleviates stalls due to WAW dependencies, it does not entirely eliminate the imbalance across rows and execution remains bottlenecked by the slowest row of PEs. To address this, each PE is provisioned with a local scratchpad buffer to temporarily hold partial sums, whether self-generated or received from upstream, in anticipation of downstream stragglers. This buffer operates as a FIFO queue, and each PE processes only the partial sums that are explicitly \emph{managed} at any given time. This can be thought of as a local context of partial sums that each PE manages.

The orchestrator actively monitors buffer occupancy, maintaining metadata to track the oldest row index present in the context queue. Upon completing MAC operations for a row, the PE flushes the corresponding oldest partial sum to its downstream neighbor (case 2.1). On the other hand, when a PE receives a partial sum (condition A), it checks whether the associated row index lies within its current execution window (condition C). If the index falls outside the active range, implying overload or skew in the upstream rows, the partial sum is temporarily bypassed (case 1.2), allowing the PE to prioritize its own computation without delay. This dynamic offloading mechanism enables heavily loaded rows to distribute surplus reduction work downstream, resulting in better utilization across the array.

Both asynchronous reduction and buffer management are realized through orchestrator microcode, which encodes the finite-state machine (FSM) logic shown in Listing~\ref{lst:spmm-fsm}. This microcode governs PE behavior by defining event-driven state transitions and generating control instructions for routing, memory, and compute operations.

The performance of {\name} on SpMM is sensitive to the scratchpad buffer capacity and the sparsity structure of the input. Larger buffers provide greater tolerance to imbalance by allowing PEs to queue more psums in its local context, thereby mitigating backpressure and reducing stall propagation. However, this comes at the cost of increased microarchitectural resources and more frequent buffer management overheads. We evaluate these trade-offs quantitatively in Section~\ref{sec:exp-spad}.

\end{shepherding}

\begin{lstlisting}[float, caption=SpMM FSM Pseudo-code, label={lst:spmm-fsm}, 
basicstyle=\tiny\ttfamily, 
keywordstyle=\bfseries, 
commentstyle=\itshape\color{gray}, 
morekeywords={msg_from_north, input, buffer, state, op, dmem_op, spad_read, spad_write, router_op,msg_to_south}]
// NNZ: Non-Zero, CID/RID: Column/Row ID of NNZ in A
msg_from_north = {None, PSum(RID)} 
input = {None, NNZ(CID), RowEnd(RID)}
buffer = (RID_start)// Buffer state
state = {MAC(RID), ACC(RID), FLUSH(RID), NOP} // FSM main State

// PE Behavior
op = MAC(CID)   if !msg_from_north && input == NNZ(CID);
     FLUSH(RID) if !msg_from_north && input == RowEnd(RID);
     ACC(RID)   if msg_from_north == PSum(RID) && 
                  buffer.is_managing(msg_from_north.RID);
     otherwise NOP;

// Memory Operations
dmem_op  = LOAD[CID] if op == MAC(CID); otherwise NOP;
spad_read  = LOAD[RID] if op == ACC(RID);
             LOAD[buffer.first()] if op == FLUSH(_) && buffer.is_full();
             otherwise NOP;
spad_write = FLUSH[RID] if op == FLUSH(RID) || (op == ACC(RID) && op.RID != state.RID);
             otherwise NOP;

// Router Behavior
router_op = NORTH_TO_SOUTH if msg_from_north == PSum(RID) && !buffer.is_managing(msg_from_north.RID) && op != FLUSH(_);
            REG_TO_SPAD if op == ACC(_) || op == FLUSH(_);
            SPAD_TO_SOUTH if op == FLUSH(_);
            SRAM_TO_REG if op == MAC(_);
            otherwise NOP;

// Buffer State Updates
buffer.RID_start += 1 if op == FLUSH(_);

// Message to south
msg_to_south = PSUM[RID] if  op == FLUSH(RID) // flush
                PSUM[RID] if op == MAC(_) && msg_from_north == PSum(RID) //bypass
\end{lstlisting}

\subsubsection{\textbf{Case Study: SDDMM}}  
Sampled Dense-Dense Matrix Multiplication (SDDMM) is a key sparse-tensor primitive, commonly employed in sparse attention mechanisms of transformer architectures~\cite{sanger, swat, vitcod, longformer, bigbird}. Conceptually, SDDMM computes \( C = M \cdot (A \times B) \), where the elementwise product with the binary mask \( M \) restricts computation to selected positions, thereby exploiting sparsity in the output.

Unlike SpMM, where sparsity appears in the input matrix, SDDMM sparsifies the output space. This makes acceleration more challenging: the arbitrary distribution of non-zeros in \( M \) induces highly irregular access and compute patterns across the array. As shown in Figure~\ref{fig:spmm-sddmm-dataflow}(b), the dataflow adheres to an inner-product style matrix multiplication, with orchestrators controlling the sparse activation of computation based on the structure of \( M \). The dense matrix \( A \) is streamed from the top edge of the PE array, while \( B \) is locally stored across PEs. At the array’s right edge, a vector of \( V \) partial products is emitted per row, selectively computed only at masked positions.

Similar to SpMM, the SDDMM dataflow also suffers from load imbalance, primarily due to the sparsity in the \( M \) matrix. PEs along the \( y \)-dimension (rows) are responsible for differing numbers of masked outputs, depending on the non-zero distribution in \( M \). This creates uneven computational demand and complicates efficient reuse of the streamed \( A \) inputs across rows.

To mitigate this, {\name} employs its scratchpad buffer in each PE, similar to SpMM, but adapted for SDDMM's needs. Here, the buffer is used to temporarily store and reuse incoming vectors from \( A \), amortizing their loading cost across multiple masked positions and balancing the compute effort among PEs. This buffering mechanism enables efficient reuse while absorbing the irregularities introduced by the mask, ultimately improving throughput. Further details of the dataflow and its implementation are provided in Appendix~\ref{sec:appendix-sddmm}.

\subsubsection{\textbf{Case study: Structured Sparsity}}
\label{sec:structured-sparsity-mapping}
Sparsity can exhibit structured patterns that are known at compile time. Common forms of structured sparsity include N:M sparsity in the input~\cite{zhou2021learning}, i.e., N non-zeros in every M elements and diagonal-wise sparsity (commonly referred to as a window) in SDDMM operations for sparse attention~\cite{longformer, bigbird, jiang2023mistral}. {\name} fully supports N:M sparsity for any N:M ratio for SpMM kernel. The global mapping is identical to SpMM, where the coordinates of sparse elements are fed to the orchestrator. With exactly N non-zero per M elements, there is no need of workload balancing with scratchpad. Instead, the psum is flushed to the next row of PEs for every N elements processed. 
The sliding window sparsity is similarly well-supported by {\name} using the mapping strategy described in prior accelerators~\cite{swat}.
Here, the output sparsity is decomposed into dense rows, where each row corresponds to a vector-matrix multiplication. The memory can be efficiently managed for perfect data reuse between the computation of rows. 

\subsection{\textbf{Mapping General Kernels}}
With our primary focus on sparse tensor operations that have been showcased previously, {\name} is able to support more general kernels. For data-agnostic kernels, i.e. where control flow remains independent of runtime data, the place-and-route-like spatial mapping techniques on reconfigurable dataflow architectures~\cite{lisa, simplemachines, riptide, cgra-me} can also be applied to our architecture as shown in Appendix~\ref{sec:canon-as-cgra}. However, as {\name} has a 4-SIMD lane, for any inner loop that cannot be unrolled in parallel by 4, such spatial mapping results in compute under-utilization. 
\begin{shepherding}
    Moreover, as orchestrators handle branching and conditionals, the associated compute must be scheduled along the PE row. Hence, inner loops within conditional branches are constrained to individual PE rows. Coupled with the time-lapse SIMD design, this imposes a lower bound on the exploitable DLP granularity for arbitrary kernels, potentially resulting in under-utilization of columns for data-dependent control loops.
\end{shepherding}

More generally speaking, {\name} can map affine loops: Let \( I \) denote the $n$-d iteration space of the loop nest:
\[
I = \{ (t_1, t_2, \ldots, s_1, s_2, \ldots) \mid t_i, s_j \in \mathbb{Z}\}, \ where
\]
\( t_1, t_2, \ldots \) are \textbf{temporal iterators}, meaning the iterations are executed sequentially with respect to the order among them.
\(s_1, s_2, \ldots \) are \textbf{spatial iterators}, meaning the iterations are executed spatially and parallely. In our case, they are only two spatial dimensions: $x$ and $y$ of the PE array.

Let \( A \) denote a memory accessed in the loop, to an $m$-d array with dimensions \( d_1, d_2, \ldots, d_m \). The \textbf{access function} \( f : \mathbb{Z}^n \to \mathbb{Z}^m \) maps the iteration space \( I \) to array indices \( A[i_1, i_2, \ldots, i_m] \). The access function \( f \) is affine for each array dimension \( i_k \):
\[
i_k = f_k(t_1, t_2, \ldots, s_1, s_2, \ldots) = c_k + \sum_{i} \beta_{ki} t_i + \sum_{j} \alpha_{kj} s_j, \ where
\]
\( c_k \in \mathbb{Z} \) is a constant offset,
\( \beta_{ki}, \alpha_{kj} \in \mathbb{Z}^2 \) are coefficients for temporal and spatial iterators, respectively.

For {\name} to be able to share data among the neighborhood PEs with the mesh-network, the spatial iterators \( s_j \) in the access function must satisfy:
\[
\exists (k,j),\alpha_{kj} \in \{-1, 0, 1\} \quad \wedge \quad  \forall (k', j') \neq (k,j), \alpha_{k'j'} = 0.
\]

\begin{shepherding}
    Although {\name} introduces a new execution model, the compiler stack is not yet fully automated. Our current implementation focuses on essential kernels (e.g., sparse/dense tensors, Polybench), and general-purpose support remains an exciting direction for future work.
\end{shepherding}

Furthermore, since finding an optimal set of loop transformations to satisfy the above conditions remains an open problem, we rely on a combination of loop analyses, primarily polyhedral-based, and with a human in the loop to select the best mapping strategy out of several candidate ones and further manually optimizing it if possible, similar to writing PTX code for GPUs~\cite{PTX}.

\section{Evaluation Methodology}

We synthesize our design as configured in Table~\ref{tab:canon-config} using a 22nm commercial FDSOI technology node and the Synopsys Design Compiler, targeting 1GHz frequency. The design incorporates a PE array with each PE featuring a 4-wide vector lane, a router, associated SRAM as data memory and a dual-ported scratchpad along with orchestrators positioned at the edge of each PE row. We further develop an event-driven, cycle-accurate simulator in \textit{Rust} to provide a detailed breakdown of performance and access patterns for every architectural component when running various workloads.

\begin{table}[h]
\centering
\caption{Configuration of the evaluated {\name} Architecture}
\renewcommand{\arraystretch}{0.75}
\resizebox{.75\linewidth}{!}{
\begin{tabular}{ll}
\toprule
\textbf{Component} & \textbf{Configuration} \\ \midrule
Array               & $8 \times 8$ 4-SIMD INT8 array;\\
SRAM              & 4KB per PE; 288KB Overall \\
Scratchpad        & Dual-port, 64 Bytes per PE \\
Orchestrator      & 8 orchestrators, 1 per PE row.\\
Main Memory       & 17 GB/s, LPDDR5x \\\bottomrule
\end{tabular}
}
\label{tab:canon-config}
\end{table}

\paragraph*{\bf Architecture.}

We evaluate {\name} against four representative baselines spanning different specialization. First, the {\bf systolic array} similar to TPU~\cite{tpu} serves as a reference for dense tensor accelerator. Next, the {\bf 2:4 systolic array} similar to NVIDIAs Tensor Core~\cite{nmsystolic, markidis2018tensorecore} for exploiting 2:4 structured sparsity, i.e., two non-zeros in every four elements. {\bf ZeD}~\cite{zed} represents the state-of-the-art specialized sparse accelerator. Finally, a {\bf conventional CGRA} is used to illustrate the general-purpose reconfigurable architecture.


To ensure fairness, all baselines are synthesized at the same technology node, with each architecture provisioned with an equal number of MAC units to guarantee equivalent theoretical peak compute performance. The CGRA baseline adopts a classical 2D-mesh PE architecture from HyCUBE~\cite{hycube} and is similar to other CGRAs~\cite{pace, pace_isocc, riptide, nowatzki2017stream-dataflow} featuring circuit-switched NoC and a small instruction memory within each PE, sufficient for mapping the most complex kernel in our benchmarks. We use a state-of-the-art CGRA mapper~\cite{morpher} for complex kernels. We develop a cycle-accurate simulator to model the timing behavior of ZeD. For a fair comparison of the architecture, we exclude ZeD's preprocessing optimization of row reorganization during evaluations, as the same can be applied to {\name}. 

On-chip memory configuration has a significant impact on power, area, and off-chip bandwidth requirements. For consistency, we allocate an average of 1KB of data memory per MAC unit for {\name} and all baseline architectures. These memories are synthesized using the foundry memory compiler at the same tech node as {\name}, with organization tailored to each architecture as shown in figure~\ref{fig:baseline_ablation}: {\name} employs distributed memory local to each PE; systolic arrays and CGRAs use memory banks along the edge of the array; and the sparse accelerator follows the original design’s bank organization~\cite{zed}. The results are validated against the respective papers to ensure fairness of comparisons.

\paragraph*{\bf Workloads}
We primarily utilize the sparse tensor kernels SpMM and SDDMM in ML workloads to demonstrate our architecture's resilience to input fragility. 
We employ the state-of-the-art sparsification technique~\cite{sparsification} to induce sparsity in the activations of the CNN and MLP layers. The inherent sparsification in the activations leads to SpMM operations. Furthermore, we apply attention sparsification techniques from~\cite{sanger, vitcod}, resulting in unstructured SDDMM operations for the QK attention matrix (later labeled SDDMM-U). We use the sliding window attention~\cite{longformer,bigbird,deepseek-nsa,jiang2023mistral}(later labeled SDDMM-Win) as a structured SDDMM operation.
These sparsification techniques enable a trade-off between sparsity and accuracy. While high levels of sparsity (> 85-90\%) may cause significant accuracy degradation depending on the model~\cite{relu-sparsity,sparsification}, we conduct experiments with sparsity levels up to 95\% to thoroughly evaluate our hardware's characteristics. 
For clarity in presenting our results, we categorize the workloads into three sparsity ranges: \textbf{S1}: Relatively dense (0–30\% sparse).  \textbf{S2}: Moderately sparse (30–60\% sparse). \textbf{S3}: Highly sparse matrices (60–95\% sparse). 

While the two sparse kernels already show the handling of different workloads, to further evaluate {\name}, we map the kernels from the PolyBenchC~\cite{karimov2018polybench} benchmark suite. Kernels of the suite containing square root or exponential operations in their loops are excluded due to the lack of support in both {\name} and CGRA. The PolyBenchC kernels are further grouped into categories based on their classification within the benchmark suite, enabling a more structured analysis. 

\section{Evaluation Results}
\begin{figure}
    \centering
      \begin{minipage}{0.45\linewidth}
        \centering
        \includegraphics[width=\linewidth]{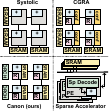}
      \end{minipage}
      \hspace{0.1em}
      \begin{minipage}{0.52\linewidth}
        \resizebox{\linewidth}{!}{
          \begin{tabular}{|c|l|l|l|}\hline
 \multicolumn{4}{|c|}{{\name} Architecture}\\\hline
          \hline
               & \makecell{\textbf{Control} \\} & \makecell{\textbf{Data} \\ }  &\makecell{\textbf{Area} \\ }\\
          \hline
               \rotatebox[origin=c]{90}{\; vs \textbf{systolic} \;} 
                   & \makecell{+Orchestrators}  
                   & \makecell{+Distributed Mem\\+Reconfig. NoC}  &+30\%\\
          \hline
               \rotatebox[origin=c]{90}{\; vs \textbf{ZeD} \;} 
                   & \makecell{-Specialized \\ Decoding\\+Orchestrators} 
                   & \makecell{-Specialized\\ Memory Banks\\-Crossbars\\+Distributed Mem\\+Reconfig. NoC}  &+9\%\\
          \hline
               \rotatebox[origin=c]{90}{\; vs \textbf{CGRA} \;} 
                   & \makecell{-Instr. Mem\\+Orchestrators} 
                   & \makecell{+Distributed Mem}  &-7\%\\
          \hline
          \end{tabular}
        }
      \end{minipage}
      \caption{Ablation of {\name}'s features through its baselines}
      \label{fig:baseline_ablation}
\end{figure}

The baselines chosen for evaluating {\name} cover a landscape of specialized and general architecture while simultaneously they function as effective ablation studies for its various features. Figure~\ref{fig:baseline_ablation} qualitatively shows key features that can be incrementally (+) added or removed (-) from the baseline architectures to ultimately yield {\name}. In the following sections, we quantify the impact of these features on resource consumption and performance, providing both a systematic, real-world ablation and valuable insights into the advantages of our architectural choices.

\subsection{Incrementally evaluating the cost of generality}

\begin{figure}
    \centering
    \includegraphics[width=.9\linewidth]{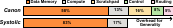}
    \caption{Area Breakdown of $\name$ and Systolic Array}
    \label{fig:area-breakdown}
\end{figure}
\begin{figure}
    \centering
    \begin{minipage}[]{0.7\linewidth}
        \centering
        \includegraphics[width=\linewidth]{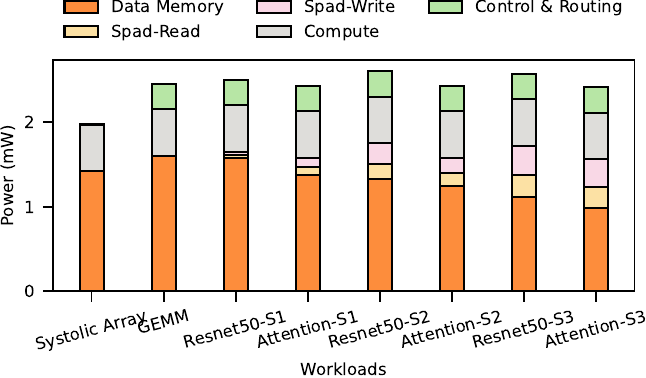}
    \end{minipage}%
    \hfill
    \begin{minipage}[]{0.3\linewidth}
        \centering
        \resizebox{\linewidth}{!}{
\begin{tabular}{|c|c|}
    \hline
    \makecell{Sparsity \\ Range} & \makecell{State \\ Transitions} \\ \hline
    S1 & 1.94e7 \\ \hline
    S2 & 3.29e7 \\ \hline
    S3 & 9.77e7 \\ \hline
\end{tabular}
    }
    \end{minipage}
    \caption{Runtime power breakdown of {\name}'s PEs (averaged), and average data-driven FSM state transitions for different sparsity ranges. Here, \textit{Spad} stands for scratchpad.}
    \label{fig:power_breakdown}
\end{figure}

\paragraph{Area cost}

A systolic array which represents the most densely packed 2D-mesh structure of compute units serves as an ablation of {\name}, wherein the routers, scratchpads, and orchestrators are omitted and the distributed memories are replaced with shared memories along the PE array edge. Figure~\ref{fig:area-breakdown} compares the area breakdown between {\name} and the systolic array, showing that {\name} incurs roughly 30\% additional area, mainly due to its scratchpads, orchestrators (control), and routing, alongside a slight increase in data-memory resources for the distributed-memory design.

Circuit-switch routers account for about 5\% of the PE chip area , enabling the mapping of more complex data dependencies than a rigid systolic NoC can support, similar to conventional CGRAs~\cite{riptide, snafu, hycube, nowatzki2017stream-dataflow}. 
For better handling irregularities, {\name} introduces orchestrators along the array edge. These orchestrators occupy 8\% of the overall area, reflecting a low control overhead. Moreover, each PE’s dual-ported scratchpad contributes 13\% to the chip area. 

Compared to the sparse accelerator, {\name} shows a 12\% of area overhead, primarily due to the added reconfiguration capabilities and generalized memory organization.
Compared to the general-purpose CGRA, benefiting from amortized control over the SIMD lane and instructions due to the orchestrators and time-lapsed execution enables {\name} to save about 7\% of total area.


\paragraph{Power cost}
Figure~\ref{fig:power_breakdown} further illustrates the power breakdown of {\name} across various workloads. Under GEMM, which employs a systolic-style dataflow, {\name} consumes nearly the same power as the systolic array, with only a slight (<13\%) overhead from control and routing. The GEMM power breakdown shows no dependency on scratchpads in regular applications. As input irregularity increases moving from low sparsity (S1) to high sparsity (S3) regimes, the FSM in each orchestrator triggers more data-driven state transitions to balance workload distribution. This necessitates using the scratchpads for buffering data to amortize the execution, resulting in a higher proportion of power dedicated to scratchpad operations and higher total power.

\textbf{Overall, {\name} incurs additional resource consumption relative to conventional domain-specific architectures; however, this trade-off enables it to achieve high performance across a wide range of workloads, as discussed in the following section.}

\subsection{Performance versus Flexibility}
\begin{figure*}
    \centering
    \includegraphics[width=0.95\linewidth]{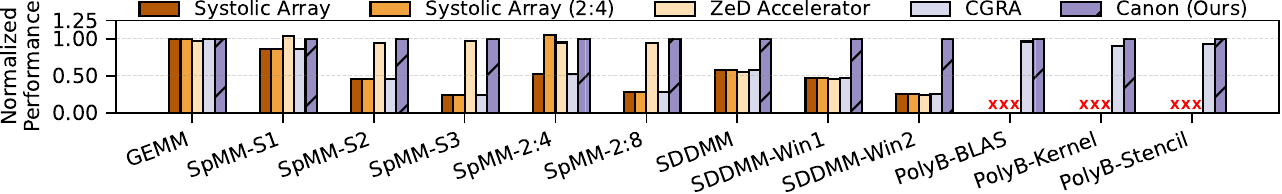}
    \caption{Speedup (fragility) of various architectures normalized to {\name} for varying kernels and input data patterns}
    \label{fig:fragility}
\end{figure*}

\begin{figure*}
    \centering
    \includegraphics[width=0.95\linewidth]{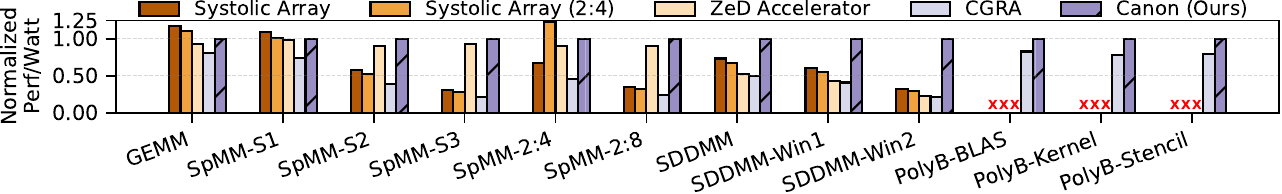}
    \caption{Performance per Watt of various architectures normalized to {\name} for varying kernels and input data patterns}
    \label{fig:ppw}
\end{figure*}

Figure~\ref{fig:fragility} and Figure~\ref{fig:ppw} quantify {\name}'s performance relative to other architectures, serving both as a study of its fragility under low performance conditions and as an ablation of certain architectural features. {\name} emulates the systolic dataflow of conventional systolic arrays for the GEMM kernel, exploiting kernel regularity to match their performance. However, as {\name} allocates a portion of its resources to generality, for extremely regular, dense workloads like GEMM, a systolic array achieves a higher performance per watt, though our results indicate this performance gap is minimal. In contrast, \textbf{when workloads exhibit sparsity that the systolic arrays dense design cannot capitalize on, their throughput can drop to less than 0.3× that of {\name}.}

When extended to support 2:4 sparse operations~\cite{tesnorcore}, the modified systolic array significantly improves its performance for the corresponding 2:4 sparse structured SpMM kernel. Nonetheless, {\bf {\name} leverages the 2:4 structure, despite being designed agnostic to it, achieving comparable performance to the modified systolic array.} Moreover, such extreme specialization does not generalize to other input patterns or kernels, as evidenced by the modified array’s diminished performance on similar (2:8 structured sparse) or dissimilar kernels (SDDDM, etc.).
{\bf In direct comparison with the sparse accelerator ZeD, {\name} demonstrates comparable performance and efficiency on unstructured sparse kernels, the very kernels for which ZeD is specialized}. ZeD outperforms marginally (<8\%) for matrices in sparsity zones S1 and select cases in S2, where specialized workload balancing through work stealing across compute units is advantageous due to a higher number of nonzeros per row. Conversely, {\name} is better at exploiting higher sparsity levels thanks to the flexibility of its scratchpad, which mitigates imbalance and provides up to 5\% better performance on some inputs. Furthermore, ZeD’s fixed datapath prevent it from leveraging structured inputs, treating all matrices as unstructured and thereby missing the additional performance gains offered in N:M sparse kernels and SDDMM-Win operations. ZeD also allocates a significant portion of its power budget to address sparsity via fully connected crossbars and specialized decoders, resulting in increased power consumption that varies with the nonzero distribution pattern. In contrast, {\name}'s homogeneous architecture opportunistically resolves load-imbalance.

SDDMM-Win1 and SDDMM-Win2 correspond to the original Longformer~\cite{longformer} configuration on BERT~\cite{BERT} (window width 512, sequence length 4K) and the Mistral-7B setup~\cite{jiang2023mistral} (window 4K, context 16K), respectively. As the other baselines architectures lack specialization for window attention, we employ the state-of-the-art sliding chunk implementation~\cite{longformer} to convert the computation into multiple dense operations. {\bf The results indicate that {\name} outperforms all baselines on window attention, with a performance gains increasing at higher sparsity (Win2) due to its ability to adapt to this input pattern.}

Finally, the CGRA, which can be deemed as a study for the importance of dynamic orchestration, must emulate the systolic dataflow for tensor operations since it has no dynamic mechanism to exploit sparsity. It delivers performance on par with systolic arrays but at the cost of higher resource consumption, as its routing and configuration circuitry is considered overprovisioned~\cite{plaid}. Nevertheless, the CGRA's design prioritizes broad programmability, enabling it to execute diverse general workloads such as those in PolyBenchC. For PolyBenchC benchmarks, CGRAs outperform {\name} in scenarios with low data parallelism, where finer-grained reconfiguration is advantageous, which constitutes some solvers in the \textit{BLAS} set. In contrast, {\bf when kernels exhibit sufficient parallelism which is typical of most other \textit{BLAS}, \textit{kernel} and \textit{stencil} workloads, {\name} achieves better performance and power efficiency.}

\begin{figure}[t]
    \centering         
    \includegraphics[width=\linewidth]{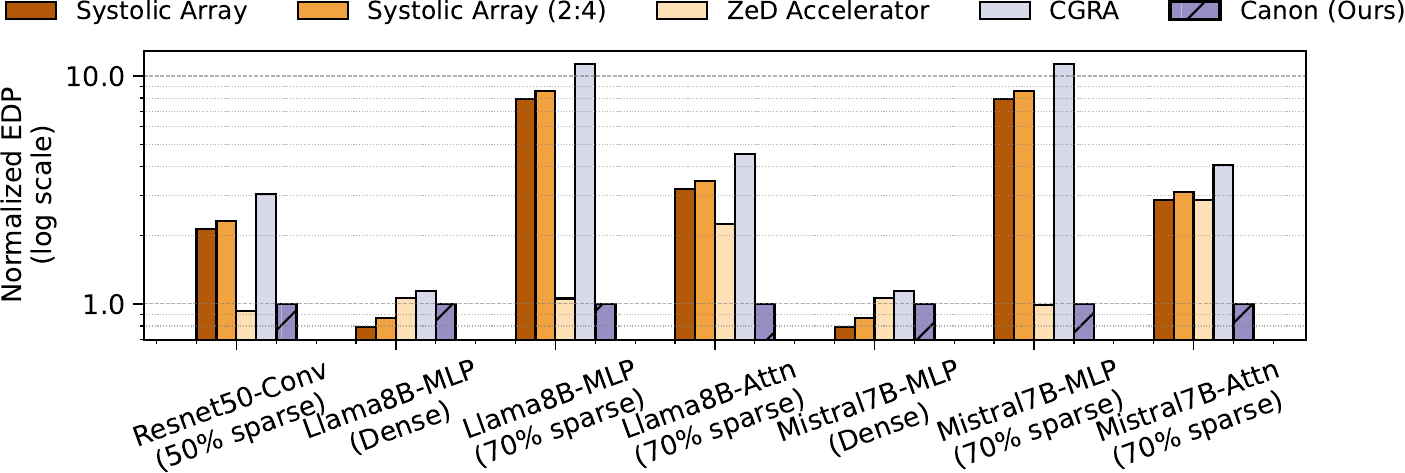}
    \caption{EDP (lower is better) of the architectures normalized to {\name} for real ML models. (in brackets: average sparsity of the model)}
    \label{fig:edp}
\end{figure}
Figure~\ref{fig:edp} presents a comparative evaluation of architectures on contemporary ML models through an analysis of the Energy-Delay Product (EDP). As previously discussed, {\name} incurs a slight efficiency overhead compared to systolic arrays for entirely dense model components. However, given the diverse composition of modern models ranging from different kernel types, such as SpMM in sparse MLPs, to a combination of SDDMM and SpMM in sparse attention, as well as varying input patterns, including unstructured sparsity in LLaMA-8B and ResNet-50 and window-structured sparsity in Mistral-7B these results highlight the critical need for a minimally fragile architecture capable of adapting to new kernels and diverse sparsity patterns. For instance, an accelerator specialized for the moderately sparse convolutions of ResNet-50 would provide little advantage when processing a diagonally windowed sparse SDDMM operation in a model like Mistral-7B.

\label{sec:exp-fragility}

\subsection{{\name}'s scalability and sensitivity to data}
\begin{figure}[t]
    \centering
    \includegraphics[width=\linewidth]{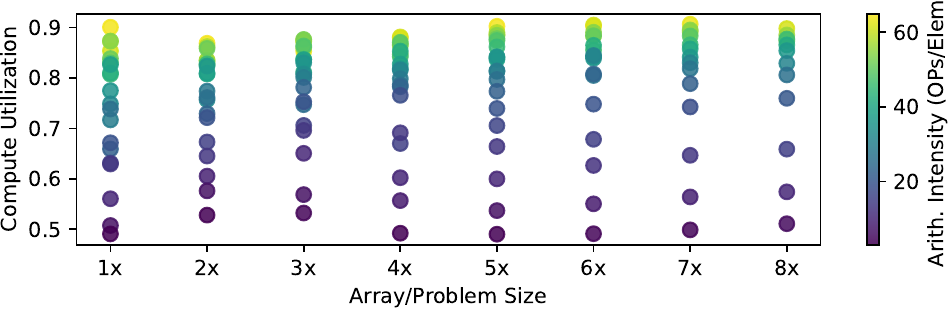}
    \caption{Sensitivity to data size \& arithmetic intensity}
    \label{fig:arith-sparsity}
\end{figure}

When operating within the compute roofline, our architecture remains entirely scalable like a systolic array. By increasing the number of PEs, we can effectively scale out our architecture. However, for the irregular workloads we target, the theoretical arithmetic intensity i.e. ``the number of computations per unit of data'' has a significant impact on performance.
This overhead occurs more frequently because each data element fetched results in only a few actual computations. To assess the sensitivity of {\name} to arithmetic intensity, we scaled both the size of the sparse tensor problems and the size of the PE array (a $8\times$ larger workload on a $8\times$ larger fabric). By adjusting the input shapes and sparsity levels, we acheive different arithmetic intensities for the inputs. Figure~\ref{fig:arith-sparsity} illustrates the compute utilization of {\name} with respect to problem size and arithmetic intensity. {\bf Our results indicate that {\name}'s compute utilization is primarily sensitive to arithmetic intensity, with no clear correlation to problem/array size, demonstrating the scalability of {\name}.}

\subsection{\textbf{Data memory Size vs. Off-Chip Bandwidth}}
The size of the on-chip data memory significantly influences the required off-chip bandwidth. Additionally, the bandwidth requirement depends on the arithmetic intensity of the workload.
To evaluate the off-chip bandwidth requirements, we randomly generate SpMM computations 
across various sparsity levels, thereby altering the arithmetic intensity. We adopt a dense-stationary tiling strategy (i.e. the dense matrix stays on-chip) because the sparse input inherently reduces off-chip traffic due to its sparsity.

\begin{figure}
    \centering
    \includegraphics[width=.9\linewidth]{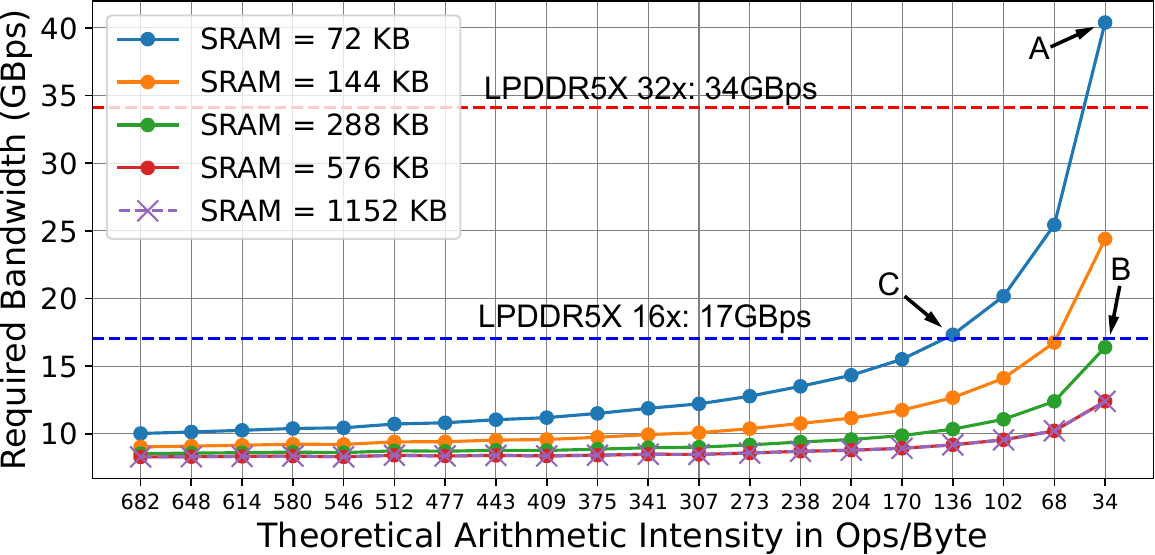}
    \caption{{\name} bandwidth requirements to hit the compute roofline for varying arithmetic intensity (sparsity increases left to right)}
    \label{fig:data-memory-sparsity}
    
\end{figure}

Figure~\ref{fig:data-memory-sparsity} presents the results. As the arithmetic intensity decreases (i.e., as sparsity increases), our architecture maintains high throughput; however, the bandwidth requirement increases because fewer computations are performed per data element. For buffer sizes large enough to hold the entire input data (SRAM size above 576KB), the off-chip bandwidth reaches its minimal value. The increased off-chip traffic at lower arithmetic intensities reflects the extra bandwidth needed for output data. It is important to note that at higher sparsity levels, although we consume more bandwidth to maintain the same amount of computation, the effective equivalent dense computation is substantially higher. For example, at a sparsity level of 95\%, we may use approximately 7x more bandwidth, but the equivalent dense throughput increases by $\approx$ 16x (not 20x due to under-utilization).

Considering the LPDDR5X as the off-chip memory, we plot the bandwidths for configurations using a single-die 16× lanes and a dual-die 32× module. The system-level design of {\name} should consider the problem's arithmetic intensity and the total cost of ownership of the on-chip and off-chip memory devices. For instance, design point A is preferable when a higher off-chip budget is available; design point B (our current configuration) is preferable when there is a higher on-chip memory budget and a higher probability of low arithmetic intensity. If the target workload is known to have higher arithmetic intensity, one can reduce the budget for both off-chip and on-chip memory (design point C).

\subsection{\textbf{Handling Load Imbalance with the Scratchpad}}
\label{sec:exp-spad}
\begin{figure}[h]
    \centering
    \includegraphics[width=\linewidth]{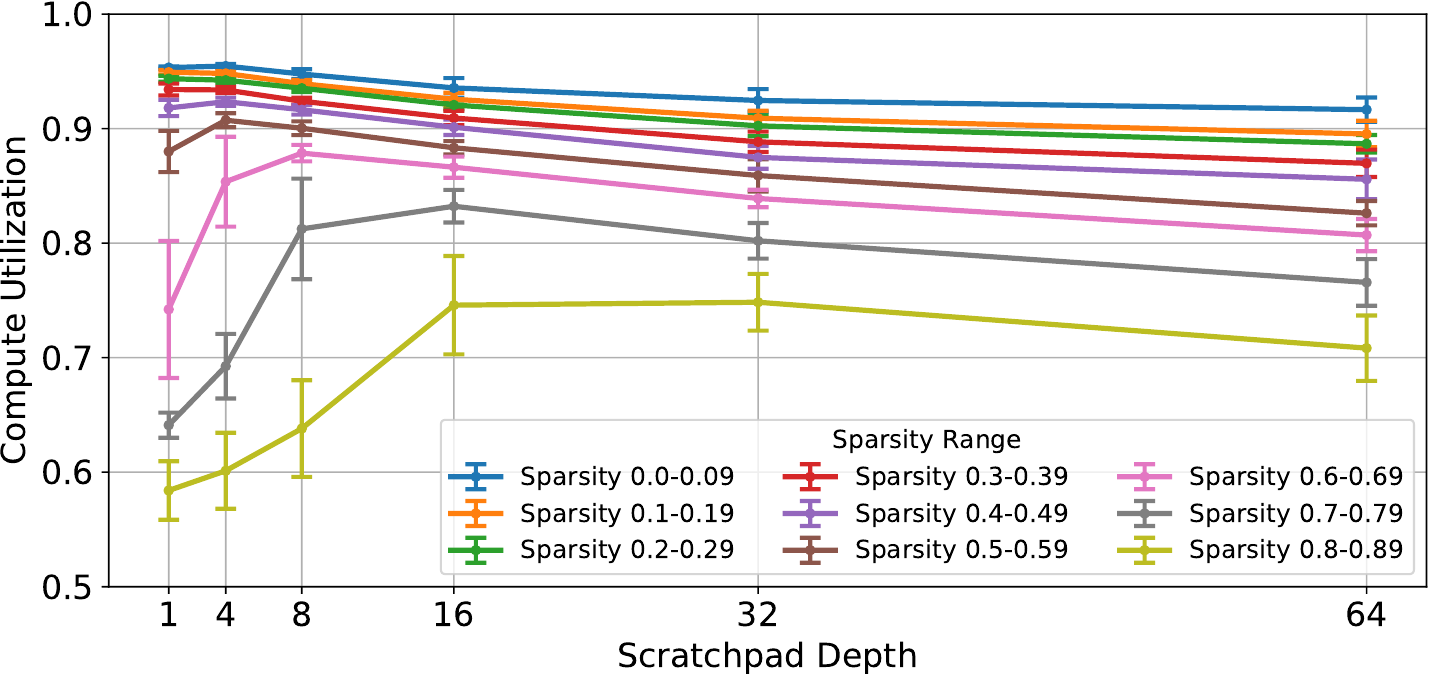}
    \caption{Impact of scratchpad depth on utilization}
    \label{fig:spmm-spad}
\end{figure}

Figure~\ref{fig:spmm-spad} presents an ablation study highlighting the impact of scratchpad size on {\name}'s performance.  An appropriately sized scratchpad can mitigate load imbalance by buffering data and amortizing the irregularity, thereby enhancing overall fabric utilization. However, if the scratchpad is too large, it introduces overhead because PEs unnecessarily buffer data in anticipation of irregularity that may not be present. As shown in the results, incorporating an optimally sized scratchpad of 16 entries leads to 10-20\% increase in compute utilization compared to 1 entry (a single register) for input sparsity levels of 60\% and above. 

Although the scratchpad size is fixed at fabrication time, the effective buffer size can be software-managed by changing the memory management logic that is programmed to the orchestrator FSM. In the performance evaluations presented in Section~\ref{sec:exp-fragility}, conservatively, we assume no prior knowledge about the sparsity and, therefore, use a buffer size of 16 entries. {\bf By incorporating compile-time knowledge about the expected sparsity range (S1, S2, S3), {\name} achieves an additional 5\% performance improvement on average by adjusting the effective scratchpad range}

\section{Related Works}



Traditional sparse accelerators hard-wire datapaths at fabrication time, fixing orchestration to a narrow set of kernels and sacrificing flexibility. A range of reconfigurable fabrics has since been proposed to better handle irregular workloads~\cite{hycube, riptide, snafu, pipestitch, nowatzki2017stream-dataflow, amber, fifer, plasticine, capstan, spu, nexus}. {\name} differs mainly in how it reconfigures and orchestrates compute: finite-state machines (FSMs) drive data-driven control over tightly coupled PEs, exploring a control/flexibility trade-off that prior designs have not. Relative to “multicore” style fabrics~\cite{dalorex, polygraph}, {\name} maintains granular computation and fine-grained inter-PE links, thereby avoiding the coarse control and data-transfer costs incurred by those systems. Compared to fine-grained fabrics like FPGAs and some CGRAs~\cite{pace, hycube}, {\name} lowers control overhead via a time-lapsed SIMD paradigm. And unlike classical CGRAs whose datapaths are fixed at compile time~\cite{pace, hycube}, {\name} supports data-driven reconfiguration at runtime.

Extending reconfigurable architectures to irregularity often brings heavy overheads~\cite{fifer}. Some works revert to multicore-style solutions~\cite{dalorex, polygraph} or enlarge PE granularity~\cite{capstan} which amortizes control complexity. Nexus Machine~\cite{nexus} uses Active-Message–based instruction flow, but retains CGRA-like configuration and resource costs. Fifer~\cite{fifer} decomposes irregular kernels into regular stages and time-multiplexes them with decoupled queues, yet pays in inter/intra-PE buffering, frequent reconfiguration, and high data-movement overheads for scheduling.
In contrast, {\name} rethinks computation on the fabric itself, abstracting control and managing irregularity through time-lapsed execution coordinated by orchestrators placed at the fabric's edge.


\textbf{Industry Architectures:}
Commercial architectures are increasingly heterogeneous, integrating specialized units to meet evolving workloads. With the rise of AI, most architectures now include systolic-array-based matrix multiplication units such as Tensor Cores in NVIDIA and AMD GPUs and Google TPUs~\cite{tpu}. Early academic projects like Plasticine~\cite{plasticine, capstan} proposed CGRAs optimized for limited parallel patterns. However, industry adoption favored embedding systolic arrays within processing elements to support dense matrix multiplication efficiently~\cite{plasticine-retrospective}, likely due to the high cost and inefficiency of reconfigurable hardware for such operations. This approach towards hardware design may reduce kernel generalizability or result in underutilized hardware, thereby negating benefits of reconfigurability~\cite{esmaeilzadeh2011dark}.
Groq~\cite{groq} employs a completely software-defined orchestration of heterogeneous components to emulate an array-level RISC-like processor. It demonstrates good performance on predictable, regular workloads yet suffers performance fragility on new applications or irregularity due to its reliance on extensive DLP and strictly compile-time mapping. In contrast, {\name}'s orchestration and compute granularity minimizes fragility while maintaining high hardware utilization for moderately DLP and irregular workloads.

\begin{shepherding}
\textbf{Understanding {\name} through a manycore lens.}
{\name} bears some resemblance to a manycore architecture. Its orchestrators are loosely analogous to the frontend of a processor (Fetch, Decode), responsible for generating control signals, while its PEs are analogous to the backend (Execute, Write-Back).
However, {\name} diverges from the conventional understanding of many-core due to its data-driven, reconfigurable orchestration. Unlike traditional scalar ISAs that require multiple compute and control instructions along with dedicated hardware to infer control decisions at runtime, {\name} shifts this burden to compile-time. The orchestration logic is statically resolved and programmed as a bitstream into the orchestrator's FSM per kernel. This enables more efficient data-dependent control flow, although the general-purpose ISA model would simplify compiler-stack integration and adoption today.

\textbf{Understanding {\name}'s time-lapsed SIMD in the context of wide SIMD architectures.}
{\name} subdivides a wide SIMD, which allows fine-grained routing between the shorter SIMD units, enabling more freedom in handling data dependencies. For instance, a row of PEs can be configured for reduction-style, pipeline-parallel execution for workloads like SDDMM. Another design consideration was that {\name} also supports a spatial execution mode (Appendix D), similar to traditional CGRA place-and-route flows, allowing static programming of distinct instructions per PE, which would not be possible with a wide SIMD. This backward compatibility was critical, given our intent to tape out a functional variant of the architecture. From a scalability standpoint, the time-lapsed SIMD model also mitigates timing closure and PnR complexity that arise when distributing instruction streams across PEs placed far from the orchestrator, especially at high frequencies.

\end{shepherding}
\section{Conclusion \& Future Work}
We present {\name}, a novel parallel architecture that integrates compile-time and runtime orchestration to overcome performance fragility across a diverse range of workloads. By employing FSM-based orchestration alongside time-lapsed SIMD execution, Canon leverages regular workload patterns such as DLP and ILP to establish a high-level dataflow while dynamically handling and adapting to irregularities. Experiments show that {\name} achieves performance on par with accelerators in their own domains, with minimal efficiency loss, all while supporting a broader spectrum of parallel applications typically accelerated by reconfigurable architectures.
Future work includes end-to-end compiler support, coming up with new techniques to exploit the architecture's programmability and dynamicity for new workloads. Canon should push the boundaries of programmable architectures, bridging the gap between ad-hoc domain specialization and general-purpose architectures.

\begin{acks}
We thank the reviewers for their feedback and interesting discussion of this work.
This research is supported by the National Research Foundation, Singapore, under its Competitive Research Program Award NRF-CRP23-2019-0003 and the Ministry of Education, Singapore, under Tier 3 grant MOE-MOET32024-0003.
\end{acks}
\clearpage
\bibliographystyle{ACM-Reference-Format}
\balance
\bibliography{references}

@String{Computing = "Computing" }

@String{Computer = "{IEEE} Computer" }

@String{Springer = "Springer-Verlag" }

@inproceedings{buffets,
  title={Buffets: An efficient and composable storage idiom for explicit decoupled data orchestration},
  author={Pellauer, Michael and Shao, Yakun Sophia and Clemons, Jason and Crago, Neal and Hegde, Kartik and Venkatesan, Rangharajan and Keckler, Stephen W and Fletcher, Christopher W and Emer, Joel},
  booktitle={Proceedings of the Twenty-Fourth International Conference on Architectural Support for Programming Languages and Operating Systems},
  pages={137--151},
  year={2019}
}

@article{relu-sparsity,
  title={Relu strikes back: Exploiting activation sparsity in large language models},
  author={Mirzadeh, Iman and Alizadeh, Keivan and Mehta, Sachin and Del Mundo, Carlo C and Tuzel, Oncel and Samei, Golnoosh and Rastegari, Mohammad and Farajtabar, Mehrdad},
  journal={arXiv preprint arXiv:2310.04564},
  year={2023}
}

@article{sparsification,
  title={Training-free activation sparsity in large language models},
  author={Liu, James and Ponnusamy, Pragaash and Cai, Tianle and Guo, Han and Kim, Yoon and Athiwaratkun, Ben},
  journal={arXiv preprint arXiv:2408.14690},
  year={2024}
}

@inproceedings{flexagon,
author = {Mu\~{n}oz-Mart\'{\i}nez, Francisco and Garg, Raveesh and Pellauer, Michael and Abell\'{a}n, Jos\'{e} L. and Acacio, Manuel E. and Krishna, Tushar},
title = {Flexagon: A Multi-dataflow Sparse-Sparse Matrix Multiplication Accelerator for Efficient DNN Processing},
year = {2023},
isbn = {9781450399180},
publisher = {Association for Computing Machinery},
address = {New York, NY, USA},
url = {https://doi.org/10.1145/3582016.3582069},
doi = {10.1145/3582016.3582069},
booktitle = {Proceedings of the 28th ACM International Conference on Architectural Support for Programming Languages and Operating Systems, Volume 3},
pages = {252–265},
numpages = {14},
location = {Vancouver, BC, Canada},
series = {ASPLOS 2023}
}

@INPROCEEDINGS{sigma,
  author={Qin, Eric and Samajdar, Ananda and Kwon, Hyoukjun and Nadella, Vineet and Srinivasan, Sudarshan and Das, Dipankar and Kaul, Bharat and Krishna, Tushar},
  booktitle={2020 IEEE International Symposium on High Performance Computer Architecture (HPCA)}, 
  title={SIGMA: A Sparse and Irregular GEMM Accelerator with Flexible Interconnects for DNN Training}, 
  year={2020},
  volume={},
  number={},
  pages={58-70},
  doi={10.1109/HPCA47549.2020.00015}}

@inproceedings{gamma,
author = {Zhang, Guowei and Attaluri, Nithya and Emer, Joel S. and Sanchez, Daniel},
title = {Gamma: leveraging Gustavson’s algorithm to accelerate sparse matrix multiplication},
year = {2021},
isbn = {9781450383172},
publisher = {Association for Computing Machinery},
address = {New York, NY, USA},
url = {https://doi.org/10.1145/3445814.3446702},
doi = {10.1145/3445814.3446702},
booktitle = {Proceedings of the 26th ACM International Conference on Architectural Support for Programming Languages and Operating Systems},
pages = {687–701},
numpages = {15},
location = {Virtual, USA},
series = {ASPLOS '21}
}

@misc{tpu,
      title={TPU v4: An Optically Reconfigurable Supercomputer for Machine Learning with Hardware Support for Embeddings}, 
      author={Norman P. Jouppi and George Kurian and Sheng Li and Peter Ma and Rahul Nagarajan and Lifeng Nai and Nishant Patil and Suvinay Subramanian and Andy Swing and Brian Towles and Cliff Young and Xiang Zhou and Zongwei Zhou and David Patterson},
      year={2023},
      eprint={2304.01433},
      archivePrefix={arXiv},
      primaryClass={cs.AR},
      url={https://arxiv.org/abs/2304.01433}, 
}

@ARTICLE{tesnorcore,
  author={Choquette, Jack and Gandhi, Wishwesh and Giroux, Olivier and Stam, Nick and Krashinsky, Ronny},
  journal={IEEE Micro}, 
  title={NVIDIA A100 Tensor Core GPU: Performance and Innovation}, 
  year={2021},
  volume={41},
  number={2},
  pages={29-35},
  doi={10.1109/MM.2021.3061394}}

@article{plasticine,
  title={Plasticine: A reconfigurable architecture for parallel paterns},
  author={Prabhakar, Raghu and Zhang, Yaqi and Koeplinger, David and Feldman, Matt and Zhao, Tian and Hadjis, Stefan and Pedram, Ardavan and Kozyrakis, Christos and Olukotun, Kunle},
  journal={ACM SIGARCH Computer Architecture News},
  volume={45},
  number={2},
  pages={389--402},
  year={2017},
  publisher={ACM New York, NY, USA}
}

@inproceedings{esmaeilzadeh2011dark,
  title={Dark silicon and the end of multicore scaling},
  author={Esmaeilzadeh, Hadi and Blem, Emily and St. Amant, Renee and Sankaralingam, Karthikeyan and Burger, Doug},
  booktitle={Proceedings of the 38th annual international symposium on Computer architecture},
  pages={365--376},
  year={2011}
}

@inproceedings{snafu,
  title={Snafu: an ultra-low-power, energy-minimal cgra-generation framework and architecture},
  author={Gobieski, Graham and Atli, Ahmet Oguz and Mai, Kenneth and Lucia, Brandon and Beckmann, Nathan},
  booktitle={2021 ACM/IEEE 48th Annual International Symposium on Computer Architecture (ISCA)},
  pages={1027--1040},
  year={2021},
  organization={IEEE}
}

@inproceedings{hycube,
  title={HyCUBE: A CGRA with reconfigurable single-cycle multi-hop interconnect},
  author={Karunaratne, Manupa and Mohite, Aditi Kulkarni and Mitra, Tulika and Peh, Li-Shiuan},
  booktitle={Proceedings of the 54th Annual Design Automation Conference 2017},
  pages={1--6},
  year={2017}
}

@article{zhou2021learning,
  title={Learning n: m fine-grained structured sparse neural networks from scratch},
  author={Zhou, Aojun and Ma, Yukun and Zhu, Junnan and Liu, Jianbo and Zhang, Zhijie and Yuan, Kun and Sun, Wenxiu and Li, Hongsheng},
  journal={arXiv preprint arXiv:2102.04010},
  year={2021}
}

@inproceedings{riptide,
  title={Riptide: A programmable, energy-minimal dataflow compiler and architecture},
  author={Gobieski, Graham and Ghosh, Souradip and Heule, Marijn and Mowry, Todd and Nowatzki, Tony and Beckmann, Nathan and Lucia, Brandon},
  booktitle={2022 55th IEEE/ACM International Symposium on Microarchitecture (MICRO)},
  pages={546--564},
  year={2022},
  organization={IEEE}
}

@article{longformer,
  title={Longformer: The long-document transformer},
  author={Beltagy, Iz and Peters, Matthew E and Cohan, Arman},
  journal={arXiv preprint arXiv:2004.05150},
  year={2020}
}

@article{bigbird,
  title={Big bird: Transformers for longer sequences},
  author={Zaheer, Manzil and Guruganesh, Guru and Dubey, Kumar Avinava and Ainslie, Joshua and Alberti, Chris and Ontanon, Santiago and Pham, Philip and Ravula, Anirudh and Wang, Qifan and Yang, Li and others},
  journal={Advances in neural information processing systems},
  volume={33},
  pages={17283--17297},
  year={2020}
}

@inproceedings{markidis2018tensorecore,
  title={Nvidia tensor core programmability, performance \& precision},
  author={Markidis, Stefano and Der Chien, Steven Wei and Laure, Erwin and Peng, Ivy Bo and Vetter, Jeffrey S},
  booktitle={2018 IEEE international parallel and distributed processing symposium workshops (IPDPSW)},
  pages={522--531},
  year={2018},
  organization={IEEE}
}

@inproceedings{karimov2018polybench,
  title={Polybench: The first benchmark for polystores},
  author={Karimov, Jeyhun and Rabl, Tilmann and Markl, Volker},
  booktitle={Technology Conference on Performance Evaluation and Benchmarking},
  pages={24--41},
  year={2018},
  organization={Springer}
}

@inproceedings{nowatzki2017stream-dataflow,
  title={Stream-dataflow acceleration},
  author={Nowatzki, Tony and Gangadhar, Vinay and Ardalani, Newsha and Sankaralingam, Karthikeyan},
  booktitle={Proceedings of the 44th Annual International Symposium on Computer Architecture},
  pages={416--429},
  year={2017}
}

@article{jiang2023mistral,
  title={Mistral 7B},
  author={Jiang, Albert Q and Sablayrolles, Alexandre and Mensch, Arthur and Bamford, Chris and Chaplot, Devendra Singh and Casas, Diego de las and Bressand, Florian and Lengyel, Gianna and Lample, Guillaume and Saulnier, Lucile and others},
  journal={arXiv preprint arXiv:2310.06825},
  year={2023}
}

@inproceedings{simplemachines,
author = {Sankaralingam, Karthikeyan and Nowatzki, Tony and Gangadhar, Vinay and Shah, Preyas and Davies, Michael and Galliher, William and Guo, Ziliang and Khare, Jitu and Vijay, Deepak and Palamuttam, Poly and Punde, Maghawan and Tan, Alex and Thiruvengadam, Vijay and Wang, Rongyi and Xu, Shunmiao},
title = {The Mozart reuse exposed dataflow processor for AI and beyond: industrial product},
year = {2022},
isbn = {9781450386104},
publisher = {Association for Computing Machinery},
address = {New York, NY, USA},
url = {https://doi.org/10.1145/3470496.3533040},
doi = {10.1145/3470496.3533040},
booktitle = {Proceedings of the 49th Annual International Symposium on Computer Architecture},
pages = {978–992},
numpages = {15},
location = {New York, New York},
series = {ISCA '22}
}

@article{gustavson,
author = {Gustavson, Fred G.},
title = {Two Fast Algorithms for Sparse Matrices: Multiplication and Permuted Transposition},
year = {1978},
issue_date = {Sept. 1978},
publisher = {Association for Computing Machinery},
address = {New York, NY, USA},
volume = {4},
number = {3},
issn = {0098-3500},
url = {https://doi.org/10.1145/355791.355796},
doi = {10.1145/355791.355796},
journal = {ACM Trans. Math. Softw.},
month = sep,
pages = {250–269},
numpages = {20}
}

@article{trips,
author = {Sankaralingam, Karthikeyan and Nagarajan, Ramadass and Liu, Haiming and Kim, Changkyu and Huh, Jaehyuk and Burger, Doug and Keckler, Stephen W. and Moore, Charles R.},
title = {Exploiting ILP, TLP, and DLP with the polymorphous TRIPS architecture},
year = {2003},
issue_date = {May 2003},
publisher = {Association for Computing Machinery},
address = {New York, NY, USA},
volume = {31},
number = {2},
issn = {0163-5964},
url = {https://doi.org/10.1145/871656.859667},
doi = {10.1145/871656.859667},
journal = {SIGARCH Comput. Archit. News},
month = may,
pages = {422–433},
numpages = {12}
}

@INPROCEEDINGS{horowitzmemory,
  author={Horowitz, Mark},
  booktitle={2014 IEEE International Solid-State Circuits Conference Digest of Technical Papers (ISSCC)}, 
  title={1.1 Computing's energy problem (and what we can do about it)}, 
  year={2014},
  volume={},
  number={},
  pages={10-14},
  doi={10.1109/ISSCC.2014.6757323}}

@inproceedings{noc,
  title={Scalable interconnects for reconfigurable spatial architectures},
  author={Zhang, Yaqi and Rucker, Alexander and Vilim, Matthew and Prabhakar, Raghu and Hwang, William and Olukotun, Kunle},
  booktitle={Proceedings of the 46th International Symposium on Computer Architecture},
  pages={615--628},
  year={2019}
}

@book{jerger2017chip,
  title={On-chip networks},
  author={Jerger, Natalie Enright and Krishna, Tushar and Peh, Li-Shiuan},
  year={2017},
  publisher={Morgan \& Claypool Publishers}
}

@inproceedings{sanger,
author = {Lu, Liqiang and Jin, Yicheng and Bi, Hangrui and Luo, Zizhang and Li, Peng and Wang, Tao and Liang, Yun},
title = {Sanger: A Co-Design Framework for Enabling Sparse Attention using Reconfigurable Architecture},
year = {2021},
isbn = {9781450385572},
publisher = {Association for Computing Machinery},
address = {New York, NY, USA},
url = {https://doi.org/10.1145/3466752.3480125},
doi = {10.1145/3466752.3480125},
booktitle = {MICRO-54: 54th Annual IEEE/ACM International Symposium on Microarchitecture},
pages = {977–991},
numpages = {15},
location = {Virtual Event, Greece},
series = {MICRO '21}
}

@inproceedings{swat,
author = {Bai, Zhenyu and Dangi, Pranav and Li, Huize and Mitra, Tulika},
title = {SWAT: Scalable and Efficient Window Attention-based Transformers Acceleration on FPGAs},
year = {2024},
isbn = {9798400706011},
publisher = {Association for Computing Machinery},
address = {New York, NY, USA},
url = {https://doi.org/10.1145/3649329.3658488},
doi = {10.1145/3649329.3658488},
booktitle = {Proceedings of the 61st ACM/IEEE Design Automation Conference},
articleno = {93},
numpages = {6},
location = {San Francisco, CA, USA},
series = {DAC '24}
}

@INPROCEEDINGS {vitcod,
author = { You, Haoran and Sun, Zhanyi and Shi, Huihong and Yu, Zhongzhi and Zhao, Yang and Zhang, Yongan and Li, Chaojian and Li, Baopu and Lin, Yingyan },
booktitle = { 2023 IEEE International Symposium on High-Performance Computer Architecture (HPCA) },
title = {{ ViTCoD: Vision Transformer Acceleration via Dedicated Algorithm and Accelerator Co-Design }},
year = {2023},
volume = {},
ISSN = {},
pages = {273-286},
doi = {10.1109/HPCA56546.2023.10071027},
url = {https://doi.ieeecomputersociety.org/10.1109/HPCA56546.2023.10071027},
publisher = {IEEE Computer Society},
address = {Los Alamitos, CA, USA},
month =mar}

@INPROCEEDINGS{lisa,
  author={Li, Zhaoying and Wu, Dan and Wijerathne, Dhananjaya and Mitra, Tulika},
  booktitle={2022 IEEE International Symposium on High-Performance Computer Architecture (HPCA)}, 
  title={LISA: Graph Neural Network based Portable Mapping on Spatial Accelerators}, 
  year={2022},
  volume={},
  number={},
  pages={444-459},
  doi={10.1109/HPCA53966.2022.00040}}

@INPROCEEDINGS{cgra-me,
  author={Chin, S. Alexander and Sakamoto, Noriaki and Rui, Allan and Zhao, Jim and Kim, Jin Hee and Hara-Azumi, Yuko and Anderson, Jason},
  booktitle={2017 IEEE 28th International Conference on Application-specific Systems, Architectures and Processors (ASAP)}, 
  title={CGRA-ME: A unified framework for CGRA modelling and exploration}, 
  year={2017},
  volume={},
  number={},
  pages={184-189},
  doi={10.1109/ASAP.2017.7995277}}

@misc{nmsystolic,
      title={Accelerating Sparse Deep Neural Networks}, 
      author={Asit Mishra and Jorge Albericio Latorre and Jeff Pool and Darko Stosic and Dusan Stosic and Ganesh Venkatesh and Chong Yu and Paulius Micikevicius},
      year={2021},
      eprint={2104.08378},
      archivePrefix={arXiv},
      primaryClass={cs.LG},
      url={https://arxiv.org/abs/2104.08378}, 
}

@inproceedings{capstan,
author = {Rucker, Alexander and Vilim, Matthew and Zhao, Tian and Zhang, Yaqi and Prabhakar, Raghu and Olukotun, Kunle},
title = {Capstan: A Vector RDA for Sparsity},
year = {2021},
isbn = {9781450385572},
publisher = {Association for Computing Machinery},
address = {New York, NY, USA},
url = {https://doi.org/10.1145/3466752.3480047},
doi = {10.1145/3466752.3480047},
booktitle = {MICRO-54: 54th Annual IEEE/ACM International Symposium on Microarchitecture},
pages = {1022–1035},
numpages = {14},
location = {Virtual Event, Greece},
series = {MICRO '21}
}

@article{plasticine-retrospective,
author = {Prabhakar, Raghu and Zhang, Yaqi and Koeplinger, David and Feldman, Matt and Zhao, Tian and Hadjis, Stefan and Pedram, Ardavan and Kozyrakis, Christos and Olukotun, Kunle},
title = {{RETROSPECTIVE}: Plasticine: A Reconfigurable Architecture For Parallel Paterns},
 booktitle = "{ISCA@50 25-Year Retrospective: 1996-2020}",
  month     = june,
  year      = 2023,
  publisher = "{ACM} {SIGARCH} and {IEEE} {TCCA}",
  url       = "https://bit.ly/isca50_retrospective",
}

@INPROCEEDINGS{amber,
  author={Carsello, Alex and Feng, Kathleen and Kong, Taeyoung and Koul, Kalhan and Liu, Qiaoyi and Melchert, Jackson and Nyengele, Gedeon and Strange, Maxwell and Zhang, Keyi and Nayak, Ankita and Setter, Jeff and Thomas, James and Sreedhar, Kavya and Chen, Po-Han and Bhagdikar, Nikhil and Myers, Zachary and D’Agostino, Brandon and Joshi, Pranil and Richardson, Stephen and Bahr, Rick and Torng, Christopher and Horowitz, Mark and Raina, Priyanka},
  booktitle={2022 IEEE Symposium on VLSI Technology and Circuits (VLSI Technology and Circuits)}, 
  title={Amber: A 367 GOPS, 538 GOPS/W 16nm SoC with a Coarse-Grained Reconfigurable Array for Flexible Acceleration of Dense Linear Algebra}, 
  year={2022},
  volume={},
  number={},
  pages={70-71},
  doi={10.1109/VLSITechnologyandCir46769.2022.9830509}}

@inproceedings{fifer,
author = {Nguyen, Quan M. and Sanchez, Daniel},
title = {Fifer: Practical Acceleration of Irregular Applications on Reconfigurable Architectures},
year = {2021},
isbn = {9781450385572},
publisher = {Association for Computing Machinery},
address = {New York, NY, USA},
url = {https://doi.org/10.1145/3466752.3480048},
doi = {10.1145/3466752.3480048},
booktitle = {MICRO-54: 54th Annual IEEE/ACM International Symposium on Microarchitecture},
pages = {1064–1077},
numpages = {14},
location = {Virtual Event, Greece},
series = {MICRO '21}
}

@inproceedings{dalorex,
   title={Dalorex: A Data-Local Program Execution and Architecture for Memory-bound Applications},
   url={http://dx.doi.org/10.1109/HPCA56546.2023.10071089},
   DOI={10.1109/hpca56546.2023.10071089},
   booktitle={2023 IEEE International Symposium on High-Performance Computer Architecture (HPCA)},
   publisher={IEEE},
   author={Orenes-Vera, Marcelo and Tureci, Esin and Wentzlaff, David and Martonosi, Margaret},
   year={2023},
   month=feb, pages={718–730} }

@INPROCEEDINGS{polygraph,
  author={Dadu, Vidushi and Liu, Sihao and Nowatzki, Tony},
  booktitle={2021 ACM/IEEE 48th Annual International Symposium on Computer Architecture (ISCA)}, 
  title={PolyGraph: Exposing the Value of Flexibility for Graph Processing Accelerators}, 
  year={2021},
  volume={},
  number={},
  pages={595-608},
  doi={10.1109/ISCA52012.2021.00053}}

@inproceedings{groq,
author = {Abts, Dennis and Ross, Jonathan and Sparling, Jonathan and Wong-VanHaren, Mark and Baker, Max and Hawkins, Tom and Bell, Andrew and Thompson, John and Kahsai, Temesghen and Kimmell, Garrin and Hwang, Jennifer and Leslie-Hurd, Rebekah and Bye, Michael and Creswick, E. R. and Boyd, Matthew and Venigalla, Mahitha and Laforge, Evan and Purdy, Jon and Kamath, Purushotham and Maheshwari, Dinesh and Beidler, Michael and Rosseel, Geert and Ahmad, Omar and Gagarin, Gleb and Czekalski, Richard and Rane, Ashay and Parmar, Sahil and Werner, Jeff and Sproch, Jim and Macias, Adrian and Kurtz, Brian},
title = {Think fast: a tensor streaming processor (TSP) for accelerating deep learning workloads},
year = {2020},
isbn = {9781728146614},
publisher = {IEEE Press},
url = {https://doi.org/10.1109/ISCA45697.2020.00023},
doi = {10.1109/ISCA45697.2020.00023},
booktitle = {Proceedings of the ACM/IEEE 47th Annual International Symposium on Computer Architecture},
pages = {145–158},
numpages = {14},
location = {Virtual Event},
series = {ISCA '20}
}

@inproceedings{pipestitch,
author = {Serafin, Nathan and Ghosh, Souradip and Desai, Harsh and Beckmann, Nathan and Lucia, Brandon},
title = {Pipestitch: An energy-minimal dataflow architecture with lightweight threads},
year = {2023},
isbn = {9798400703294},
publisher = {Association for Computing Machinery},
address = {New York, NY, USA},
url = {https://doi.org/10.1145/3613424.3614283},
doi = {10.1145/3613424.3614283},
booktitle = {Proceedings of the 56th Annual IEEE/ACM International Symposium on Microarchitecture},
pages = {1409–1422},
numpages = {14},
location = {Toronto, ON, Canada},
series = {MICRO '23}
}

@article{deepseek-v3,
  title={Deepseek-v3 technical report},
  author={Liu, Aixin and Feng, Bei and Xue, Bing and Wang, Bingxuan and Wu, Bochao and Lu, Chengda and Zhao, Chenggang and Deng, Chengqi and Zhang, Chenyu and Ruan, Chong and others},
  journal={arXiv preprint arXiv:2412.19437},
  year={2024}
}

@article{deepseek-nsa,
  title={Native Sparse Attention: Hardware-Aligned and Natively Trainable Sparse Attention},
  author={Yuan, Jingyang and Gao, Huazuo and Dai, Damai and Luo, Junyu and Zhao, Liang and Zhang, Zhengyan and Xie, Zhenda and Wei, YX and Wang, Lean and Xiao, Zhiping and others},
  journal={arXiv preprint arXiv:2502.11089},
  year={2025}
}

@ARTICLE{specialization-question,
  author={Nowatzki, Tony and Gangadhar, Vinay and Sankaralingam, Karthikeyan and Wright, Greg},
  journal={IEEE Micro}, 
  title={Domain Specialization Is Generally Unnecessary for Accelerators}, 
  year={2017},
  volume={37},
  number={3},
  pages={40-50},
  doi={10.1109/MM.2017.60}}

@inproceedings{zed,
  title={Zed: A generalized accelerator for variably sparse matrix computations in ml},
  author={Dangi, Pranav and Bai, Zhenyu and Juneja, Rohan and Wijerathne, Dhananjaya and Mitra, Tulika},
  booktitle={Proceedings of the 2024 International Conference on Parallel Architectures and Compilation Techniques},
  pages={246--257},
  year={2024}
}

@article{plaid,
  title={Enhancing CGRA Efficiency Through Aligned Compute and Communication Provisioning},
  author={Li, Zhaoying and Yin, Chenyang and Bandara, Thilini Kaushalya and Juneja, Rohan and Tan, Cheng and Bai, Zhenyu and Mitra, Tulika},
  journal={arXiv preprint arXiv:2412.08137},
  year={2024}
}

@article{BERT,
  title={Roberta: A robustly optimized bert pretraining approach},
  author={Liu, Yinhan and Ott, Myle and Goyal, Naman and Du, Jingfei and Joshi, Mandar and Chen, Danqi and Levy, Omer and Lewis, Mike and Zettlemoyer, Luke and Stoyanov, Veselin},
  journal={arXiv preprint arXiv:1907.11692},
  year={2019}
}

@inproceedings{morpher,
  title={Morpher: An open-source integrated compilation and simulation framework for cgra},
  author={Wijerathne, Dhananjaya and Li, Zhaoying and Karunaratne, Manupa and Peh, Li-Shiuan and Mitra, Tulika},
  booktitle={Fifth Workshop on Open-Source EDA Technology (WOSET)},
  year={2022}
}

@inproceedings{diag,
author = {Wang, Dong Kai and Kim, Nam Sung},
title = {DiAG: a dataflow-inspired architecture for general-purpose processors},
year = {2021},
isbn = {9781450383172},
publisher = {Association for Computing Machinery},
address = {New York, NY, USA},
url = {https://doi.org/10.1145/3445814.3446703},
doi = {10.1145/3445814.3446703},
booktitle = {Proceedings of the 26th ACM International Conference on Architectural Support for Programming Languages and Operating Systems},
pages = {93–106},
numpages = {14},
location = {Virtual, USA},
series = {ASPLOS '21}
}

@ARTICLE{mit-dataflow,
  author={Arvind and Nikhil, R.S.},
  journal={IEEE Transactions on Computers}, 
  title={Executing a program on the MIT tagged-token dataflow architecture}, 
  year={1990},
  volume={39},
  number={3},
  pages={300-318},
  doi={10.1109/12.48862}}

@inproceedings{protobuf,
author = {Karandikar, Sagar and Leary, Chris and Kennelly, Chris and Zhao, Jerry and Parimi, Dinesh and Nikolic, Borivoje and Asanovic, Krste and Ranganathan, Parthasarathy},
title = {A Hardware Accelerator for Protocol Buffers},
year = {2021},
isbn = {9781450385572},
publisher = {Association for Computing Machinery},
address = {New York, NY, USA},
url = {https://doi.org/10.1145/3466752.3480051},
doi = {10.1145/3466752.3480051},
booktitle = {MICRO-54: 54th Annual IEEE/ACM International Symposium on Microarchitecture},
pages = {462–478},
numpages = {17},
location = {Virtual Event, Greece},
series = {MICRO '21}
}

@inproceedings{scalable_interconnects,
author = {Zhang, Yaqi and Rucker, Alexander and Vilim, Matthew and Prabhakar, Raghu and Hwang, William and Olukotun, Kunle},
title = {Scalable interconnects for reconfigurable spatial architectures},
year = {2019},
isbn = {9781450366694},
publisher = {Association for Computing Machinery},
address = {New York, NY, USA},
url = {https://doi.org/10.1145/3307650.3322249},
doi = {10.1145/3307650.3322249},
booktitle = {Proceedings of the 46th International Symposium on Computer Architecture},
pages = {615–628},
numpages = {14},
location = {Phoenix, Arizona},
series = {ISCA '19}
}

@article{PTX,
  title={An In-depth Study on the Performance Impact of CUDA, OpenCL, and PTX Code},
  author={Memarzia, Puya and Khunjush, Farshad},
  journal={J. Inf. Comput. Sci},
  volume={10},
  number={2},
  pages={124--136},
  year={2015},
  publisher={Citeseer}
}

@ARTICLE{eyeriss,
  author={Chen, Yu-Hsin and Krishna, Tushar and Emer, Joel S. and Sze, Vivienne},
  journal={IEEE Journal of Solid-State Circuits}, 
  title={Eyeriss: An Energy-Efficient Reconfigurable Accelerator for Deep Convolutional Neural Networks}, 
  year={2017},
  volume={52},
  number={1},
  pages={127-138},
  doi={10.1109/JSSC.2016.2616357}}

@inproceedings{spu,
author = {Dadu, Vidushi and Weng, Jian and Liu, Sihao and Nowatzki, Tony},
title = {Towards General Purpose Acceleration by Exploiting Common Data-Dependence Forms},
year = {2019},
isbn = {9781450369381},
publisher = {Association for Computing Machinery},
address = {New York, NY, USA},
url = {https://doi.org/10.1145/3352460.3358276},
doi = {10.1145/3352460.3358276},
booktitle = {Proceedings of the 52nd Annual IEEE/ACM International Symposium on Microarchitecture},
pages = {924–939},
numpages = {16},
location = {Columbus, OH, USA},
series = {MICRO '52}
}

@misc{nexus,
      title={Nexus Machine: An Active Message Inspired Reconfigurable Architecture for Irregular Workloads}, 
      author={Rohan Juneja and Pranav Dangi and Thilini Kaushalya Bandara and Tulika Mitra and Li-shiuan Peh},
      year={2025},
      eprint={2502.12380},
      archivePrefix={arXiv},
      primaryClass={cs.AR},
      url={https://arxiv.org/abs/2502.12380}, 
}

@inproceedings{pace,
  title={PACE: A scalable and energy efficient CGRA in a RISC-V SoC for edge computing applications},
  author={Nambiar, Vishnu P and Chong, Yi Sheng and Bandara, Thilini Kaushalya and Wijerathne, Dhananjaya and Li, Zhaoying and Juneja, Rohan and Peh, Li-Shiuan and Mitra, Tulika and Do, Anh Tuan},
  booktitle={2024 IEEE Hot Chips 36 Symposium (HCS)},
  pages={1--1},
  year={2024},
  organization={IEEE}
}

@INPROCEEDINGS{pace_isocc,
  author={Nambiar, V. P. and Chong, Y. S. and Bandara, T. K. and Wijerathne, D. and Li, Z. and Juneja, R. and Peh, L.-S. and Mitra, T. and Do, A. T.},
  booktitle={2024 21st International SoC Design Conference (ISOCC)}, 
  title={A 360 GOPS/W CGRA in a RISC-V SoC with Multi-Hop Routers and Idle-State Instructions for Edge Computing Applications}, 
  year={2024},
  volume={},
  number={},
  pages={89-90},
  doi={10.1109/ISOCC62682.2024.10762131}}

\appendix
\newpage
\section{Detailed SpMM Mapping} Listing~\ref{lst:spmm-code} illustrates the high-level dataflow for SpMM. This listing is obtained from transforming the conventional matrix multiplication depicted in Listing~\ref{lst:gemm} by applying loop tiling, reordering, and splitting. The loop is splitted into two parallel tasks: the first, spanning lines 13–22, involves the PEs receiving sparse inputs and executing MAC operations to generate local partial sums, while the second, spanning lines 25–28, handles the asynchronous accumulation of these partial sums, as discussed in Section~\ref{sec:spmm}.
\begin{lstlisting}[caption=GeMM original code,label={lst:gemm},basicstyle=\scriptsize\ttfamily]
A[M][K]
B[K][N]
C[M][N]
fn gemm(T* A, T* B, T* C){
 for m in 0..M
  for n in 0..N
   for k in 0..k
    C[m][n] += A[m][k] * B[k][n]
}
\end{lstlisting}

\begin{figure}[h]
    \centering
    \includegraphics[width=.95\linewidth]{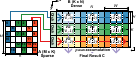}
    \caption{SpMM dataflow}
    \label{fig:spmm-dataflow-appendix}
\end{figure}

\begin{lstlisting}[caption=SpMM Dataflow Pseudo-code (transformed),label={lst:spmm-code},basicstyle=\scriptsize\ttfamily]
// See Figure 17 for these hyper-parameters
W,H: width, height of the tile of each PE
M,K,N: Matrix Shapes
X,Y: PE array dimensions
ASSERT W * X = N; 
ASSERT  Y * H = K;
// Memory Layout
A[M][Y][H]: tiled input sparse matrix A // A[M][K]
B[X][Y][W][H]: tiled input dense matrix B // B[K][N]
C[M][X][W]: output matrix C // C[M][N]
psum[M][X][Y][w]; // intermediate psums produced by PE(X, Y)

fn spmatmul(T* A, T* B, T* C){ 
 // local psum computation
 for m in 0..M // sequential execution  (temporal)
  for h in 0..H // sequential execution (temporal)
   parallel_for x in 0..X // parallel execution  (spatial)
    if (A[m][y][h] != 0) {// executed by orchestrators
     parallel_for y in 0..Y // parallel execution (spatial)
     // scalar-vector multiplication and accumulation
      _sv_mac(psum[m][x][y], A[m][y][h], B[x][y][h]);
    } // End if non-zero

 // accumulation of partial sums (asynchronous, non-associative)
 for (a = 0: a < Y; a++) // 
  for (b = 0; b < X; b++) // 
   for (m = 0; m < M; m++) // 
    _vv_acc(psum[m][a][b], C[m][a]);
}
\end{lstlisting}

\section{Detailed SDDMM Mapping}
\label{sec:appendix-sddmm}
Listing~\ref{lst:sddmm-code} illustrates the SDDMM dataflow. In each cycle, a vector from matrix \(A\) with width \(V\) (corresponding to the vector lane size) is streamed into the first PE from the top. Simultaneously, a bitmask \(M\) is provided to the orchestrator to select the column of \(B\) that will be dot-multiplied with the incoming \(A\) vector. Each PE then executes element-wise MAC operations between the \(A\) vector and the corresponding locally stored \(B\) vector.
Each PE processes \(H\) columns of \(B\); these columns are partitioned into \(\frac{H}{V}\) cycles of vectored MAC operations. This process reduces the accumulation dimension from \(H\) to \(V\), yielding a partial sum vector of width \(V\). These partial sums are propagated from left to right and are accumulated in a \(V\) dimension vector. Finally, the reduction of the \(V\)-dimensional partial sum to a single scalar is performed by the last column of PEs, just before the result is forwarded to the memory controllers.
\begin{figure}[t]
    \centering
    \includegraphics[width=.58\linewidth]{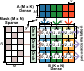}
    \caption{SDDMM dataflow}
    \label{fig:sddmm-dataflow-appendix}
\end{figure}

\begin{lstlisting}[caption=SDDMM Dataflow Pseudo-code ,label={lst:sddmm-code},basicstyle=\scriptsize\ttfamily]
W, H: width, height of the tile of each PE
M,K,N: Matrix Shapes
X,Y: PE array dimensions, V: Vector lane width
ASSERT W * X * V = K
ASSERT  Y * H = N;
// Memory Layout
A[M][W][X][V]: tiled input sparse matrix A // A[M][K]
B[X][Y][H][W][V]: tiled input dense matrix B // B[K][N]
C[M][Y][H]: output matrix C // C[M][N]
M[M][Y][H]: the mask of the result matrix // M[M][N]
psum[X][Y][V]
psum_reduced[Y][V]
psum_final[Y]

fn sddmm(T* A, T* B, T* C, T* M){
 for m in 0..M // sequential execution (temporal)
  for h in 0..H // sequential execution (temporal)
   for w in 0..W // sequential execution (temporal)
    parallel_for x in 0..X // parallel execution (spatial)
     parallel_for y in 0..Y // parallel execution (spatial)
      if (M[m][y][h] != 0){
       _vv_mac(psum[x][y], A[m][w][x], B[x][y][h][w])
      }
 for y in 0..Y
  for x in 0..X
   // reduce psum from shape (x,y,v) to (y,v)
   _vv_acc(psum_reduced[y],psum[x][y])
 for y in 0..Y
  for v in 0..V
   // reduce to the final psum
   psum_final[y] = _sum(psum_reduced[y])
}
\end{lstlisting}

\section{FSM behavioural insights}

\begin{figure}[b] \centering \includegraphics[width=.9\linewidth]{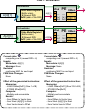} \caption{SpMM detailed execution: case 1} \label{fig:spmm-detailed-1} \end{figure}

\begin{figure}[b] \centering \includegraphics[width=.9\linewidth]{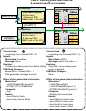} \caption{SpMM detailed execution: case 2} \label{fig:spmm-detailed-2} \end{figure}

\begin{figure}[h] \centering \includegraphics[width=\linewidth]{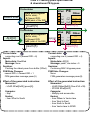} \caption{SpMM detailed execution: case 3} \label{fig:spmm-detailed-3} \end{figure}

Figure~\ref{fig:spmm-detailed-1},~\ref{fig:spmm-detailed-2}, and~\ref{fig:spmm-detailed-3} show three representative snapshots of the SpMM kernel execution. For illustration purpose, we use the same dense matrix B for mapping as in Figure~\ref{fig:spmm-dataflow-appendix}, where each PE stores only two rows of B. For clarity, the figures display only the orchestrator’s internal buffer-management state and the first PE of each row, focusing on the top two rows of the PE array to illustrate their interaction. To simplify, we consider a scratchpad of size 4.

\paragraph{Case 1 (Normal Operation)} When a new non-zero element arrives at the orchestrator’s input and no message is received from a neighboring orchestrator, the orchestrator issues a MAC operation for its row of PEs. It calculates the address in the local memory (i.e., a slice of B) based on the column index of the incoming element from A, ensuring that each PE fetches the correct data from its local memory. For instance, if the input metadata in the first row indicates a non-zero element from A at row 4, column 1 (case \textcircled{1}, A[4][1]), the local memory address that the PE needs to load from the corresponding B matrix is computed as (1 modulo 2), since two rows of B are mapped to each row of PE.

 \paragraph{Case 2 (Row Completion and Psum Forwarding)} Once a row finishes processing, indicated by a row-end token, space in the scratchpad must be freed for a new row. According to our FIFO buffer management policy, the oldest buffered partial sum (psum) is flushed to the downstream PE. The state meta registers record the row ID (RID) of the oldest element and its corresponding offset in the scratchpad (to implement a circular FIFO).

The orchestrator also informs the downstream orchestrator about the row index of the forwarded psum (e.g., psum(1) for the psum of row index 1\footnote{psum(1)represents the inter-orchestrator message, while psum[1] denotes the actual value.}). Upon receiving this message, the downstream orchestrator checks whether the row index of the incoming psum falls within its current range of responsibility (based on the index of the newest buffered row and the buffer length). If it does (as in case 2), the downstream row stops its local psum computation and accumulates the incoming psum into its scratchpad.

\paragraph{Case 3 (Early psum Arrival and Bypass)} If the downstream orchestrator receives a psum whose row index falls outside its current range—in this case, receiving psum 5 while the newest row being computed is row 2—it bypasses the received psum without interrupting its local computation. This situation indicates a workload imbalance, as the downstream orchestrator is “too late,” potentially due to an excessive number of sparse elements in its previous row. The bypass is achieved using the router by directly forwarding the psum from north to south without interrupting the execution pipeline. The downstream orchestrator also notifies its next downstream orchestrator about the bypassed psum.

\begin{figure*}[h]
    \centering
    \includegraphics[width=\linewidth]{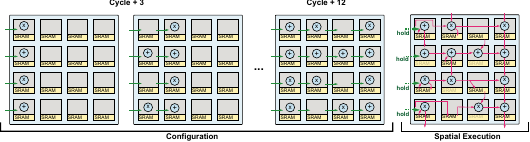}
    \caption{Using {\name} for spatial dataflow execution}
    \label{fig:pure-spatial}
\end{figure*}

\section{Using Canon for Spatial Execution} \label{sec:canon-as-cgra} {\name} inherently can support a fully static place-and-route style spatial mapping, in which each PE executes the same instruction over time. This mapping is typically associated with entirely spatially reconfigurable architectures where a kernel's dataflow graph can be entirely mimicked on the fabric. Figure~\ref{fig:pure-spatial} illustrates how this is accomplished. During the configuration phase, the orchestrator preloads instructions into the PE array without immediate execution—the results are discarded. For instance, fully configuring a 4-column PE array requires 12 cycles (4 columns × 3 cycles). During the execution phase, PEs can be set to a “hold” state to prevent instructions from propagating further through the pipeline to subsequent PEs. This effectively means stopping the \textit{staggered instruction issue}. This mechanism allows the PEs to execute the preloaded instruction from the configuration phase, with the orchestrators maintaining the hold signal to ensure that the PEs remain in this state.
\end{document}